\documentclass[10pt,aps,prd,floatfix,showpacs,twocolumn,superscriptaddress]{revtex4-2}

\usepackage[normalem]{ulem}
\usepackage{graphicx}
\usepackage{amsmath, amsfonts, amssymb, bm}
\usepackage{soul}
\usepackage{ulem}
\usepackage{tabularx}
\usepackage{slashed}

\usepackage{amstext}
\usepackage{mathrsfs}
\usepackage{hyperref}
\usepackage{color}
\usepackage{slashed}

\begin{document}

\title{Enhancement of nonlinear pair production in a flying-focus pulse}
\author{Md Reshad Ur Rahman}
\email{mrahm18@ur.rochester.edu}
\affiliation{Department of Physics and Astronomy, University of Rochester, Rochester, New York 14627, USA}
\affiliation{Laboratory for Laser Energetics, University of Rochester, Rochester, New York 14623, USA}
\author{Martin S. Formanek}
\affiliation{ELI Beamlines Facility, The Extreme Light Infrastructure ERIC, 252 41 Doln\'{i} B\v{r}e\v{z}any, Czech Republic}
\author{Elias Gerstmayr}
\affiliation{School of Mathematics and Physics, Queen's University Belfast, BT7 1NN, Belfast, UK}
\author{Dillon Ramsey}
\affiliation{Laboratory for Laser Energetics, University of Rochester, Rochester, New York 14623, USA}
\author{John P. Palastro}
\affiliation{Laboratory for Laser Energetics, University of Rochester, Rochester, New York 14623, USA}
\author{Antonino Di Piazza}
\email{a.dipiazza@rochester.edu}
\affiliation{Department of Physics and Astronomy, University of Rochester, Rochester, New York 14627, USA}
\affiliation{Laboratory for Laser Energetics, University of Rochester, Rochester, New York 14623, USA}

\begin{abstract}
A photon can decay into an electron-positron pair in the presence of an intense laser pulse by a process known as nonlinear Breit-Wheeler pair production (NBWPP). For a sufficiently intense pulse, the probability of NBWPP for a high-energy gamma photon scales more favorably with the interaction time than with the field intensity. In contrast to typical stationary-focus Gaussian (SFG) laser pulses, a flying-focus (FF) pulse features a focal point that moves at a programmable velocity. Here, we show that a FF pulse counterpropagating with a high-energy photon beam while its focus copropagates with the beam at the speed of light enhances the NBWPP yield relative to an equal-energy SFG pulse. Numerical simulations show that a Joule-class tightly-focused FF pulse allows for an enhancement of the pair-production yield by 12\%, 29\% and 76\% for a photon of 10, 20 and 50 GeV energy, respectively. Thus, in this regime, FF pulses are more efficient at producing electron-positron pairs than conventional SFG pulses.
\end{abstract}

\maketitle

Strong-field quantum electrodynamics (SFQED) studies processes occurring in electromagnetic background fields comparable to the Schwinger critical field $F_{\text{cr}}=m^2/|e|=1.3\times10^{16}~\text{V/cm}=4.4\times 10^{13}~\text{G}$, corresponding to a peak laser intensity of $I_{\text{cr}}=4.6 \times 10^{29}~\text{W/cm}^2$ \cite{Ritus:1985vta, Baier_b_1998,DiPiazza:2011tq,Gonoskov:2021hwf,Fedotov:2022ely}. Here, $m$ and $e<0$ are the electron mass and charge, respectively, and we use natural units $\hbar=c=\epsilon_0=1$ throughout, so that $\alpha=e^2/4\pi\approx 1/137$ is the fine-structure constant.

Since the invention of chirped pulse amplification \cite{Strickland_1985}, high-intensity laser technology has advanced significantly \cite{Danson2019}. The highest peak intensity achieved to date is $\sim10^{23}~\text{W/cm}^2$ \cite{Yoon:2021ony}. At such intensities, the peak electromagnetic field experienced by a multi-GeV electron in its rest frame exceeds $F_{\text{cr}}$. This has motivated numerous investigations of two basic SFQED processes in strong background laser fields involving the lightest charged particles, namely electrons and positrons \cite{Ritus:1985vta, Baier_b_1998,DiPiazza:2011tq,Gonoskov:2021hwf,Fedotov:2022ely}: nonlinear Compton scattering (NCS), i.e., the emission of a high-energy photon by an electron or positron, and nonlinear Breit-Wheeler pair production (NBWPP), i.e., the decay of a photon into an electron-positron pair. NCS, along with the related phenomenon of radiation reaction, has already been experimentally demonstrated \cite{E144:1996enr, Cole_2018, Poder_2018, Mirzaie_2024, Los_2026}. Typical configurations to detect NBWPP consist of a laser pulse colliding with a high-energy electron beam, with the resulting NCS photons decaying into pairs (trident pair production) \cite{Burke:1997ew}. In the initial experimental demonstrations reported in Refs.~\cite{E144:1996enr,Burke:1997ew}, the available laser intensity was modest ($\sim 10^{18}
~\text{W/cm$^2$}$) but was compensated by the extremely high electron-beam energy, of order $\sim 50$~GeV. Several proposed experimental programs at high-intensity laser facilities aim to further study radiation and pair production in the strong-field regime \cite{Weber2017, Gales2018, Yoon:2021ony, Nees2020, NSF_OPAL}. Moreover, producing electron-positron pairs using high-power laser pulses remains of interest not only from a fundamental perspective but also as a tool for laboratory astrophysics \cite{Vranic2018, He2018}.

In contrast to the stationary focus of laser pulses typically used in SFQED experiments, flying-focus (FF) laser pulses feature a focal point that moves at a programmable velocity \cite{Sainte-Marie_2017, Froula_2018, Turbull2018}. The moving focus allows the peak-intensity location to propagate over distances much longer than the Rayleigh length. Techniques to produce FF pulses include chromatic optics \cite{Froula_2018, Turbull2018, Jolly:2020}, nonlinear optical processes \cite{simpson2020nonlinear, simpson2022spatiotemporal}, axiparabolas paired with echelons \cite{Palastro:2020gcl, pigeon2024ultrabroadband}, and plasma optics \cite{Li_2024}. 
Prior applications of FF pulses to SFQED have been proposed to study NCS and radiation reaction \cite{DiPiazza:2020wxp,Formanek:2023mkx, Formanek2025}. The structure of FF fields also enables analytical studies of SFQED phenomena, including field depletion \cite{Adamo_2025}.

\begin{figure}[]
	\begin{center}
	\includegraphics[width=\linewidth]{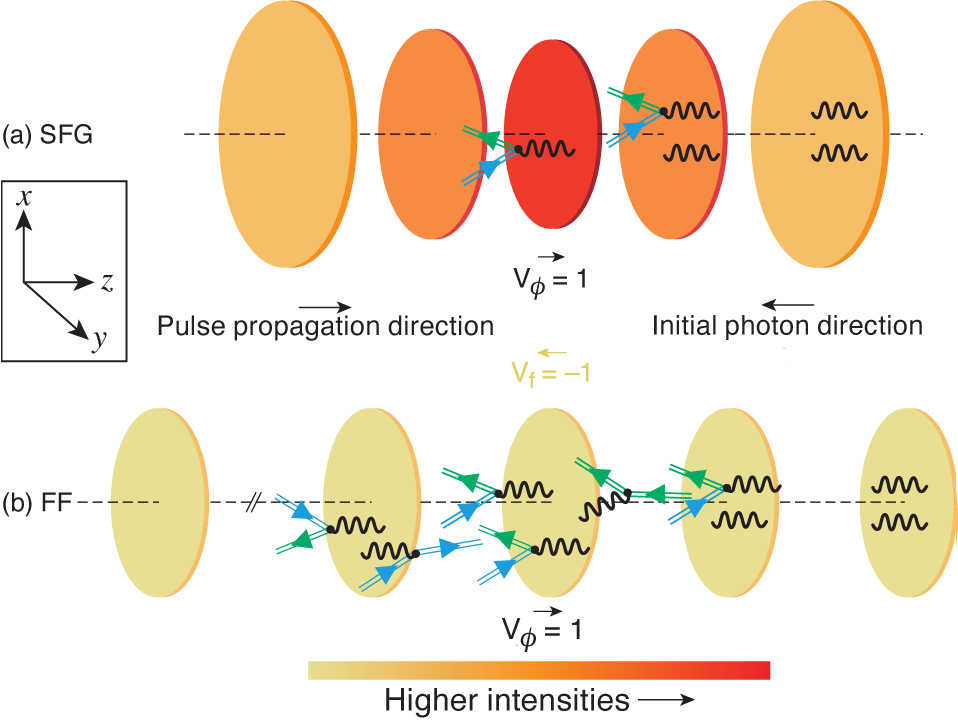}
           \vspace{-0.8cm}
	\end{center}
	\caption{Schematic setup of NBWPP in a head-on collision between a high-energy photon beam (wavy lines) and either a SFG pulse [part (a)] or a FF pulse [part (b)] of equal energy. The peak-intensity location of the SFG pulse is fixed, whereas that of the FF pulse counterpropagate with respect to its phase velocity at the speed of light, i.e., the focal velocity is $v_f=-1$ while the phase velocity is $v_{\phi}=1$ along the $z$ direction. The line break in the FF pulse emphasizes that it is much longer than the SFG pulse, implying that at a fixed laser energy the SFG pulse is much more intense at focus than the FF pulse.} 
    \label{Schematic}
\end{figure}

In this letter, we show that FF pulses can enhance the electron-positron pair yield from NBWPP relative to stationary-focused Gaussian (SFG) pulses of equal energy. This enhancement is obtained in the schematic shown in Fig.~\ref{Schematic}, where high-energy photons collide head-on with either a SFG pulse or a FF pulse, with the FF focus comoving with the incoming photons at the speed of light.
In the regime of interest here, the pair-production yield scales more favorably with pulse length than with pulse intensity. As a result, the extended interaction enabled by the FF allows for more pairs, despite the lower intensity of the FF pulse compared to the SFG. Furthermore, we will see that the produced pairs emit high-energy photons via NCS, which approximately counterpropagate relative to both the FF and SFG pulses but remain close to the moving focus of the FF pulse. These photons may subsequently decay into additional electron–positron pairs, giving rise to a shower-like cascade, which is more effective for a FF than for an equal-energy SFG.

The configuration requires a 1-J FF pulse with an intensity of $10^{20}\;\text{W/cm$^2$}$, the latter exceeding the current record by an order of magnitude \cite{Liberman2026PRR, Liberman2026FFLWFA,Arrowsmith_2026}, along with a multi-GeV photon beam. While, Joule-class flying-focus pulses have been already demonstrated \cite{Kabacinski_2023}, the high-energy photons can be produced via inverse Compton scattering (ICS) of high-energy electrons, such as those produced at SLAC (approximately 50~GeV) \cite{E144:1996enr,Burke:1997ew} and at the Large Electron-Positron (LEP) collider at CERN (approximately 100~GeV) \cite{LEP}. At present, electron beams with energies of order 10~GeV are available both at large conventional accelerators \cite{Facet-II, LUXE:2024TDR} and at high-power laser facilities using laser-wakefield acceleration \cite{Picksley_2024}. Planned laser facilities such as the NSF OPAL are expected to accelerate electrons to energies of order 100~GeV \cite{Shaw2025OPAL100GeV}. Finally, three future large electron-positron colliders are currently under discussion: the Future Circular Collider (FCC) \cite{FCC} and the Compact Linear Collider (CLIC) \cite{CLIC} at CERN, as well as the International Linear Collider (ILC) in Japan \cite{ILC}. These machines plan to reach energies beyond 200~GeV and up to 1~TeV. The relatively modest laser energy required for FF-enhanced NBWPP makes it feasible to co-locate the laser system also with such large-scale facilities.

We characterize SFQED processes using the classical and quantum nonlinearity parameters \cite{DiPiazza:2011tq, Burton:2014wsa, Gonoskov:2021hwf, Fedotov:2022ely}. The classical nonlinearity parameter $\xi_0=|e|E_0/m\omega_0$ quantifies the dimensionless field strength, where $E_0$ is the field amplitude and $\omega_0$ is the central angular frequency. For a photon, the quantum nonlinearity parameter is $\kappa=|e|\sqrt{-(F^{\mu\nu}k_\nu)^2}/m^3$, where $F^{\mu\nu}$ is the laser-field electromagnetic tensor, $k^\nu=(\omega,\bm{k})$ is the photon four-momentum and the metric tensor is $\eta^{\mu\nu}=\text{diag}(+1,-1,-1,-1)$. Below, we denote the local value of the photon quantum nonlinearity parameter in the field as $\kappa$, whereas $\kappa_0$ indicates its peak value. Below, we assume $\xi_0\gg 1$, as routinely achieved at existing high-power laser facilities, and $\xi_0^3\gg\kappa_0$, which allows locally-constant field approximation (LCFA) \cite{Ritus:1985vta, Baier_b_1998, DiPiazza:2011tq, Gonoskov:2021hwf, Fedotov:2022ely}. The LCFA is particularly useful here because it enables to obtain the probability of a SFQED process such as NBWPP in an arbitrary field from the corresponding result in a constant crossed field for which analytical expressions are available \cite{Ritus:1985vta, Baier_b_1998, DiPiazza:2011tq, Gonoskov:2021hwf, Fedotov:2022ely}. We recall that a constant crossed field is a field in which the electric and magnetic fields are constant, orthogonal, and equal in magnitude.

Within the LCFA, the NBWPP rate is approximated to an accuracy of $3\%$ by \cite{Baier_b_1998}
\begin{equation} \label{interp}
\begin{split}
    \frac{dP}{dt}&\approx \left(\frac{3}{8}\right)^{3/2}\frac{\alpha}{2}\frac{m^2}{\omega} \kappa e^{-8/(3\kappa)}\\
    &\times \left[(1+C_1\kappa)^{-1/3}+(1+C_2\kappa)^{-1/3}\right],
\end{split}
\end{equation}
where $C_1=0.094$ and $C_2=0.94$, and $\kappa$ is the local value of the photon quantum nonlinearity parameter. Thus, for $\kappa\ll 1$ the NBWPP probability is exponentially suppressed with decreasing laser-field amplitude, whereas the scaling changes significantly when $\kappa$ is of order unity or larger. In the complementary regime $\kappa\gg 1$, the rate scales as $\kappa^{2/3}$ (see Eq.~\eqref{interp} and also Refs.~\cite{Ritus:1985vta, Baier_b_1998, DiPiazza:2011tq, Gonoskov:2021hwf, Fedotov:2022ely}).

We consider the scheme shown in Fig.~\ref{Schematic}, in which a conventional stationary-focused Gaussian (SFG) pulse [part (a)] or a FF pulse [part (b)] of equal energy counterpropagates with respect to a high-energy photon beam. In the FF case, the focus moves at the speed of light in the same direction as the photons, allowing them to remain in the region of highest field over distances much longer than the Rayleigh length. In contrast, the SFG pulse cannot remain tightly focused over the full pulse length when that length exceeds its Rayleigh length. If a photon counterpropagates with respect to a linearly polarized FF pulse whose focus copropagates with the photon at the speed of light, it experiences an oscillating field with approximately constant amplitude such that the local quantum nonlinearity parameter is $\kappa_{\text{FF}}(t)=\kappa_0\sin(\Psi_0)$, where $\kappa_0=2\xi_0\omega_0\omega/m^2$ and $\Psi_0$ is the time-dependent phase (see Eq.~(S8) of the Supplemental Material (SM) \cite{SM}). Assuming that at the maximum of each oscillation satisfies $\kappa_0\gg 1$ and that the FF duration is much longer than a single oscillation period, the scaling of the pair-production probability can be inferred by integrating the corresponding asymptotic expression in Eq.~\eqref{interp} over time:
\begin{equation} \label{est}
\begin{split}
    P &\approx 0.39 \frac{\tau[\text{fs}]}{\omega[\text{GeV}]} \kappa_0^{2/3}=0.63 \left(\frac{U[\text{J}]}{\omega[\text{GeV}]}\frac{\tau^2[\text{fs}]}{\sigma^2_0[\mu\text{m}]}\right)^{1/3},
    \end{split}
\end{equation}
where $\tau$, $U$, $\sigma_0$ are the duration, energy, and the waist size of the FF pulse, respectively (strictly speaking, the large-$\kappa$ asymptotic does not hold at all times for an oscillating field, but this can be neglected for the purpose of the estimate in Eq.~\eqref{est}). Considering a photon bunch with transverse size much smaller than $\sigma_0$ and longitudinal size much smaller than the Rayleigh length $\eta_0=\omega_0\sigma_0^2$ of the FF pulse, $P$ estimates the number of pairs $\hat{N}_\pm$ produced per incoming photon, provided that $P<1$ (note that a photon crossing the entire FF beam interacts with the field for a time $\tau/2$). Equation~\eqref{est} implies $P\propto \tau I_0^{1/3}$, where $I_0$ is the peak laser intensity, and therefore shows a more favorable scaling of the pair-production yield with interaction time than with field intensity. Equivalently, for two pulses with the same energy and waist, the pair-production yield is higher for the longer pulse. For an SFG pulse, $P$ is smaller than Eq.~\ref{est} for the same pulse parameters because (within leading-order paraxial theory) $\kappa_{\text{SFG}}(t)=\kappa_0\sin(\Psi_0)/\sqrt{1+t^2/z_0^2}$, with $z_0=\omega_0\sigma_0^2/2$ being the Rayleigh length of the SFG pulse and with the replacements $\eta\rightarrow z$ and $\eta_0\rightarrow z_0$ in $\Psi_0$ (see Eq.~(S8) of the SM \cite{SM}). Thus, compared to an equal-energy SFG pulse we expect an enhancement of the pair yield when the FF pulse length is much longer than the Rayleigh length of the SFG pulse.

A fraction of the photons decay via NBWPP into a first generation of electron-positron pairs. Especially for the longer interaction time of the FF pulse, these first-generation particles have a greater probability to undergo NCS \cite{Formanek2025}, producing photons that can decay into a second generation of pairs. Eventually, for higher-generation photons $\kappa_0$ decreases to values for which pair production becomes negligible. Notably, the FF focus effectively copropagates not only with the primary photons but also with the produced pairs and higher-generation photons. This occurs because we consider photon energies $\omega\gg m\xi_0$ and only a few generations of SFQED processes develop, so that if $\varepsilon$ is the incoming-particle energy in either NCS or NBWPP, the typical emission angle relative to the incoming-particle direction is $m\xi_0/\varepsilon\ll 1$ \cite{Ritus:1985vta, Baier_b_1998, DiPiazza:2011tq, Gonoskov:2021hwf, Fedotov:2022ely}. This implies that the particles have small transverse momenta and are traveling near/at the speed of light, so they can remain in the high-intensity of the FF pulse over an extended distance. Provided that higher-generation photons have sufficiently large $\kappa$, they also contribute to the enhancement of the pair-production yield relative to the SFG pulse. These qualitative expectations are confirmed numerically below.

We investigate NBWPP in both FF and SFG pulses using Ptarmigan \cite{Blackburn2023}, which simulates laser-pulse-particle interactions (including NCS and NBWPP) using a Monte Carlo method within the LCFA. The Monte Carlo algorithm uses the known probabilities of NCS and NBWPP in a constant crossed field \cite{Ritus:1985vta, Baier_b_1998, DiPiazza:2011tq, Gonoskov:2021hwf, Fedotov:2022ely}, which, as noted above, approximate the local probabilities in an arbitrary field as long as $\xi_0\gg 1$ and $\xi_0^3\gg\kappa_0$ \cite{Ritus:1985vta, Reiss:1962nhe, Baier_b_1998,DiPiazza:2018bfu,Ilderton_2019_b,Di_Piazza_2019}. We have verified that these conditions are satisfied in all simulations. To compare SFG and FF pulses, we have implemented a module in Ptarmigan to compute the fields of a FF pulse. The implementation details together with the analytical expressions of the FF electromagnetic field are provided in the SM \cite{SM}.

Ptarmigan requires $\xi_0$, which is obtained from the average power $\mathcal{P}_{\text{avg}}$. Specifically, $U=\mathcal{P}_{\text{avg}}\tau$ and $\mathcal{P}_{\text{avg}}=\frac{\pi}{4}(m/e)^2\xi_0^2\sigma_0^2\omega_0^2$ for both the SFG and FF pulses, which yields a value of $\xi_0$ that is reasonably accurate even for tightly focused pulses (see the SM \cite{SM}).

We consider three photon beams, each consisting of 10{,}000 photons propagating along the negative $z$ axis (the laser pulses propagate along the positive $z$ axis; see Fig.~\ref{Schematic}). The beams differ only in photon energy, with $\omega=10$, 20, and 50~GeV, respectively. The primary photons are initialized with no spatial or momentum spread and, for the FF pulse, are co-located with the center of the moving focus. 

We simulated laser-photon interactions using 1-J and 5-J laser pulses and the photon beams described above. Unless otherwise stated, all pulses are tightly focused to a waist $\sigma_0=1.01\lambda_0$, with central wavelength $\lambda_0=2\pi/\omega_0$ set to $0.8~\mu$m. This tight focusing was selected for both the FF and the SFG pulse by scanning the parameter space at fixed laser energy, as it yields the largest pair yield $\hat{N}_{\pm}$ per incoming photon (see also below). The corresponding Rayleigh lengths are $\eta_0=5.1~\mu\text{m}=17~$fs for the FF pulse and $z_0=2.6~\mu\text{m}=8.6~$fs for the SFG pulse. To accurately account for the tight focusing of the SFG pulse, the simulations include up to the fourth-order correction to the paraxial approximation \cite{Salamin_2007}. For the FF pulse, the fields reported in the SM \cite{SM} are exact in the monochromatic limit for arbitrary waist size. Finally, we consider pulse durations $\tau>20~$fs~$\sim\eta_0$ that are available at high-power laser facilities \cite{DiPiazza:2011tq, Gonoskov:2021hwf, Fedotov:2022ely}. Under these conditions and based on the scalings for $P$ presented above, noticeable differences in the pair-production yield $\hat{N}_\pm$ between SFG and FF pulses are expected when $\tau>\eta_0$.

\begin{figure}[]
	\begin{center}
		\includegraphics[width=\linewidth]{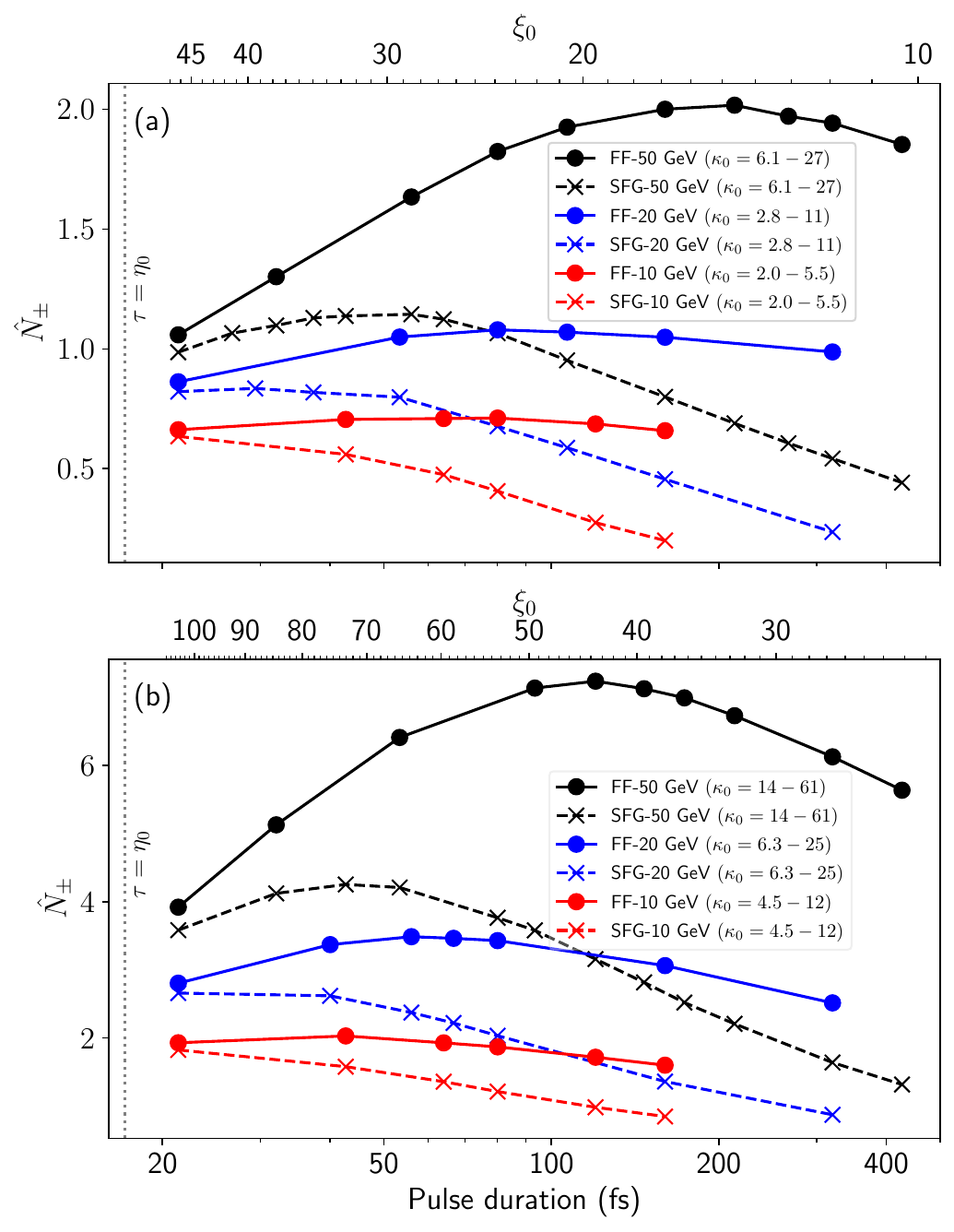}
	\end{center}
       \vspace{-0.8cm}
	\caption{Number of electron-positron pairs $\hat{N}_\pm$ produced per incoming photon for tightly focused FF and SFG pulses with energies of (a) 1-J and (b) 5-J in head-on collisions with photons of energy $\omega=10$, 20, and 50~GeV, respectively. The corresponding ranges of the values of $\kappa_0$ are indicated in the legend. The peaks in $\hat{N}_\pm$ occur, for increasing photon energies at $\kappa_0=2.8$, 6.9, 8.7 (FF) and $\kappa_0=5.5$, 9.4, 17 (SFG) for 1~J pulses and $\kappa_0=12$, 15, 26 (FF), and at $\kappa_0=12$, 25, 43 (SFG) for 5~J pulses.} 
    \label{Comparison1&5J}
\end{figure}

The results for the number of pairs $\hat{N}_\pm$ produced per incoming photon for the 1-J pulse are shown in Fig.~\ref{Comparison1&5J}(a).
The FF pulse provides a significant advantage over the SFG pulse when its peak $\hat{N}_\pm$ occurs at a pulse length that is much larger than $\eta_0$. At their respective maxima, the FF pulse produces $12\%$, $29\%$, and $76\%$ more pairs for photon energies of 10, 20, and 50~GeV, respectively. This advantage arises because the primary photons, as well as the photons produced in subsequent generations via NCS, satisfy $\kappa_0\gg1$ for $\tau\gg\eta_0$, where the NBWPP probability scales more favorably with interaction time than with field intensity (see Eq.~\eqref{est}). The enhancement increases with higher photon energies because $\kappa_0$ increases. However, the higher-generation photons produced via NCS (which also contribute to pair production) have lower energies than the primary photons, making it more difficult to maintain $\kappa_0\gg1$ over many generations.

Increasing the pulse energy to 5~J increases the probability that a given photon decays earlier than in the 1-J case. This shifts the peaks of $\hat{N}_\pm$ for a given photon energy toward shorter pulse lengths (see Fig.~\ref{Comparison1&5J}(b)). However, the FF peaks still occur at pulse lengths much longer than $\eta_0$, yielding an advantage over the SFG pulse of $11\%$, $31\%$, and $70\%$ for photon energies of 10, 20, and 50~GeV, respectively (to be compared with the 1-J case). In both Figs.~\ref{Comparison1&5J}(a) and \ref{Comparison1&5J}(b), one also finds that the pair yield $\hat{N}_\pm$ per incoming photon decreases much more slowly for FF pulses than for SFG pulses once $\tau$ exceeds the optimal pulse length. Thus, even at lower laser intensities, provided that $\kappa_0\gg 1$ is satisfied, FF pulses would still outperform a standard SFG pulse.

\begin{figure}[]
	\begin{center}
		\includegraphics[width=\linewidth]{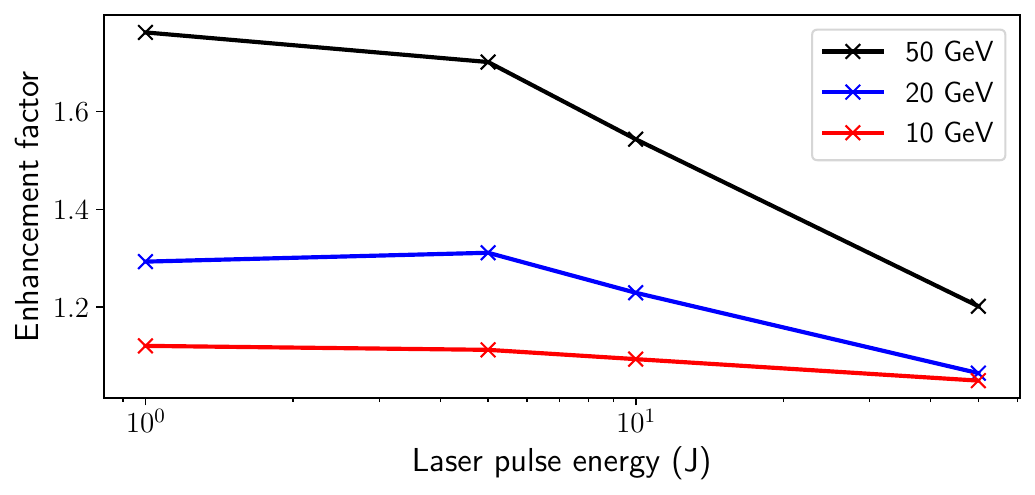}
	\end{center}
        \vspace{-0.8cm}
	\caption{Enhancement factor, defined as the ratio of the maximum number of electron-positron pairs produced by the FF pulse to that produced by the SFG pulse at a given photon energy, as a function of laser pulse energy. The ratio tends to decrease with increasing pulse energy because the optimal FF pulse lengths become comparable to $\eta_0$.} 
    \label{Enhancement}
\end{figure}

To quantitatively compare the pair yield for the two laser pulses, we introduce the enhancement factor, defined as the ratio of the maximum $\hat{N}_\pm$ produced by the FF pulse to that produced by the SFG pulse. Figure~\ref{Enhancement} shows the enhancement factor for four laser energies. The enhancement decreases with the laser energy because at higher pulse energies the optimal FF pulse lengths become comparable to the Rayleigh length $\eta_0$, making the two pulse configurations essentially equivalent.

We have already mentioned that showers must occur in the considered laser-photon collisions, in the sense that photons emitted by the created electrons and positrons subsequently decay into additional pairs. Indeed, Fig. \ref{Comparison1&5J} shows that $\hat{N}_\pm>1$ for both laser energies. Under the conditions considered here, showers are generated predominantly through incoherent channels, where at each formation length only a single elementary process (NCS and/or NBWPP) occurs. Coherent processes are negligible at the parameters discussed in this letter compared to incoherent processes \cite{Di_Piazza_2010, DiPiazza:2011tq, Gonoskov:2021hwf, Fedotov:2022ely}. This is consistent with Ptarmigan, which includes only incoherent channels in the shower development.

It is instructive to compare the properties of the showers produced by FF and SFG pulses. At the point of maximum $\hat{N}_\pm$ in Fig.~\ref{Comparison1&5J} (corresponding to $\omega=50$~GeV), the FF produces predominantly first- and second-generation pairs, accounting for $48\%$ and $49\%$, respectively, with third-generation pairs remaining rare ($2.3\%$). The FF pulse produces a percent-level larger number of second-generation pairs than first-generation pairs, even though the photons producing them have lower energies than the primary photon. This occurs because almost all electrons and positrons from the first generation radiate a high-energy photon that can subsequently decay. In contrast, for the SFG pulse and the same photon beam ($\omega=50$~GeV), the pairs are mostly first generation ($66\%$), followed by second generation ($33\%$), with third generation remaining rare ($1.2\%$). Thus, the FF pulse produces more pairs in each generation than the SFG pulse, consistent with its ability to maintain an approximately constant field amplitude over distances much longer than the SFG Rayleigh length. In the 5-J case (see Fig.~\ref{Comparison1&5J}(b)), the percentages of pairs at $\omega=50$~GeV in the first four generations are $14\%$, $56\%$, $27\%$, and $3.1\%$ for the FF pulse and $22\%$, $59\%$, $17\%$, and $1.2\%$ for the SFG pulse, respectively. At this higher laser pulse energy, the second-generation fraction exceeding the first-generation fraction also occurs for the SFG pulse. For completeness, the SM \cite{SM} reports the energy distribution of the decayed photons in each generation for the 1-J and 5-J pulses at the parameters yielding the maximum $\hat{N}_\pm$.
\begin{figure}[]
	\begin{center}
		\includegraphics[width=\linewidth]{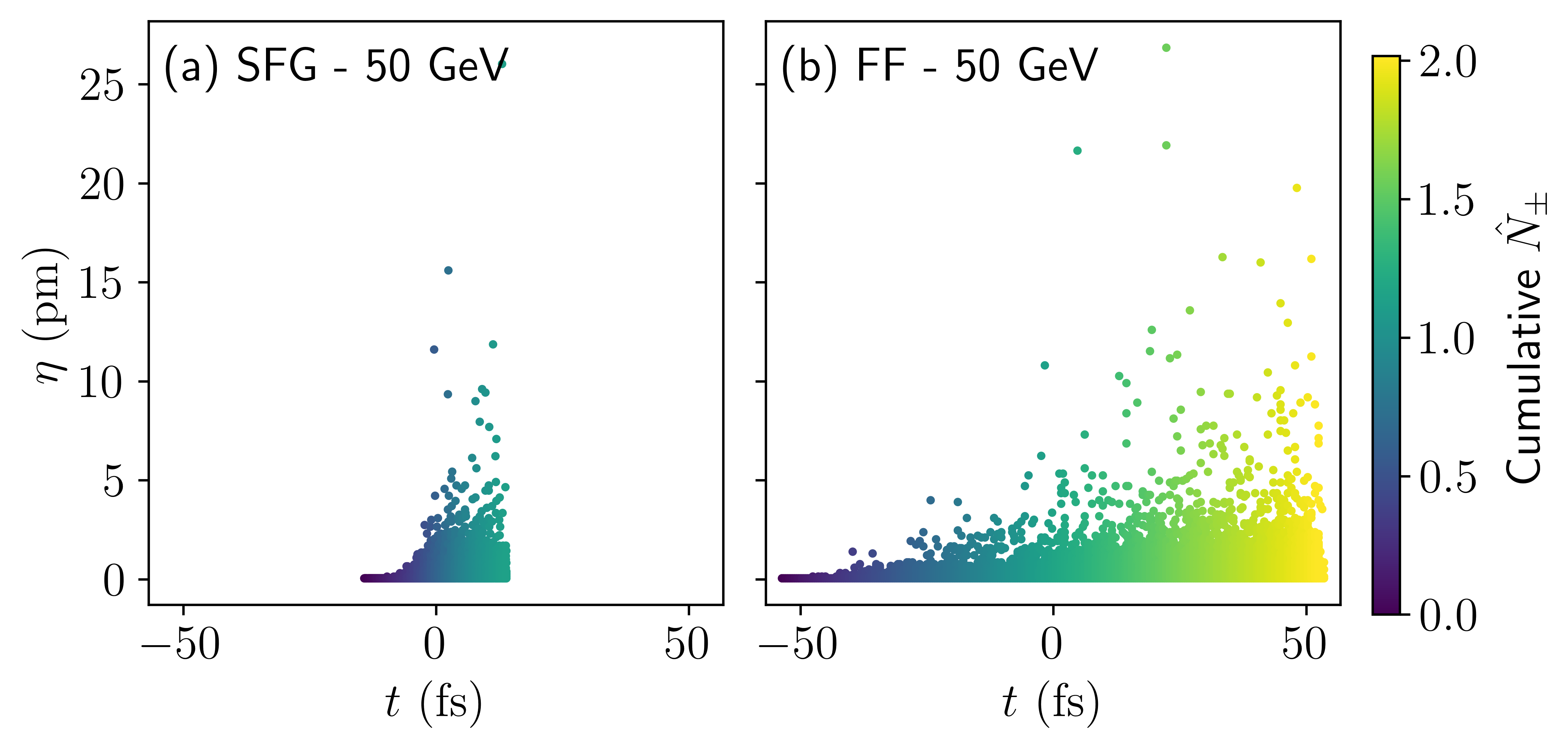}
	\end{center}
       \vspace{-0.8cm}
	\caption{Spacetime locations where pairs are produced for (a) the SFG pulse and (b) the FF pulse, using a photon beam with $\omega=50~$GeV. Each point represents a small spacetime region in which pairs are produced, and the color indicates the cumulative number of pairs per incoming photon, $\hat{N}_\pm$, produced up to that point in the $(\eta,t)$ plane, where $\eta=t+z$. The pulse parameters that yield the maximum $\hat{N}_\pm$ for a pulse energy of 1~J are used.} 
    \label{Spacetime}
\end{figure}

The difference in the production mechanism for the 1-J FF and SFG pulses that produce most pairs is illustrated in Fig.~\ref{Spacetime}. The figure shows the cumulative number of produced pairs as a function of time and the light-cone variable $\eta=t+z$, where each point marks a small spacetime region in which pairs are produced and the color indicates the cumulative $\hat{N}_\pm$ up to that point in the $(\eta,t)$ plane. One can see that pairs are produced throughout the FF interaction along $z\approx -t$, whereas for the SFG pulse most pairs are produced close to $t=z=0$, consistent with the discussion above. The electrons and/or positrons in the primary pairs can still have $\chi\gg1$, where $\chi=|e|\sqrt{-(F^{\mu\nu}p_\nu)^2}/m^3$ is the quantum nonlinearity parameter of a charged particle and $p^\nu$ is its four-momentum. Such particles also have a high probability of emitting high-energy photons in a longer equal-energy FF pulse \cite{Formanek2025}. These higher-generation photons continue to move with the FF focus and may also decay into pairs. As a result, pair production occurs over a wider time interval in the FF pulse than in the SFG pulse. For both laser pulses, higher-generation pairs have $\sim 1~$mrad-scale divergence, leading to $\eta\lesssim10^{-5}\lambda_0$. Note that for all primary photons $\eta=t+z=0$, and any subsequently produced pairs/photons can only lag behind the moving focus, such that their $\eta>0$.

\begin{figure}[]
	\begin{center}
		\includegraphics[width=\linewidth]{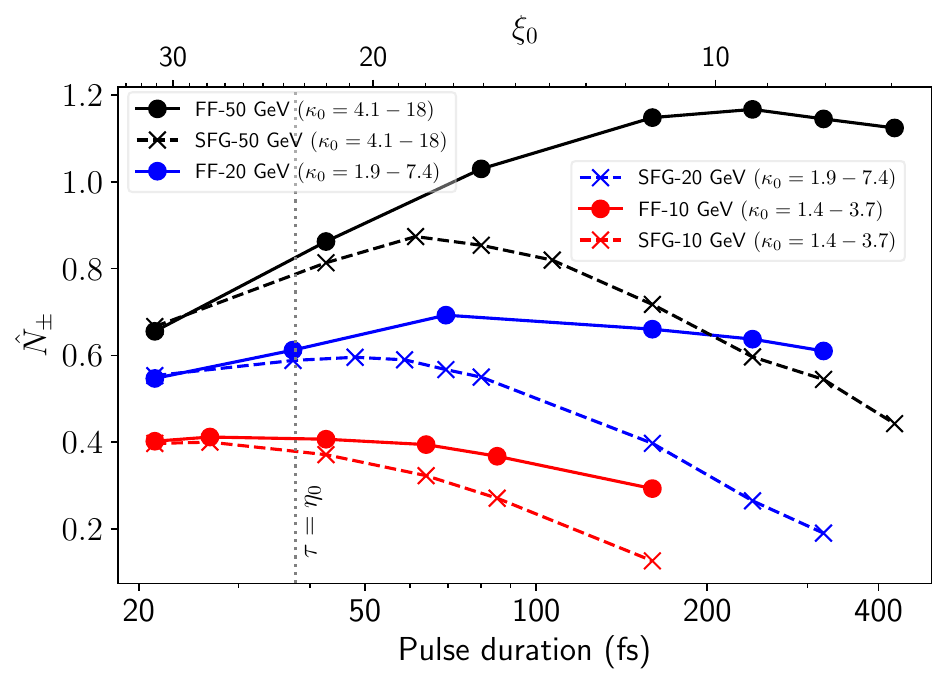}
	\end{center}
       \vspace{-0.8cm}
	\caption{Number of electron-positron pairs $\hat{N}_\pm$ produced per incoming photon for 1-J FF and SFG pulses colliding with photons of energy $\omega=10$, 20, and 50~GeV, respectively. Both pulses have $\sigma_0=1.5\lambda_0$. The advantage of the FF focus is reduced compared to Fig.~\ref{Comparison1&5J}(a) because the maximum pair yield for the FF pulse occurs at a pulse length that is not much larger than the increased $\eta_0$.} 
    \label{larger waist}
\end{figure}

As we have mentioned, the highest values of $\hat{N}_\pm$ for both SFG and FF pulses are achieved with tight focusing. In this case, most of the available laser energy goes into increasing the field intensity and extending the pulse length. Figure~\ref{larger waist} shows explicitly that increasing the waist of both the FF and SFG pulses (while keeping the laser energy at 1~J) reduces $\hat{N}_\pm$ in both cases.
The FF advantage decreases to $3.0\%$, $16\%$, and $34\%$ for photon energies of $\omega=10$, 20, and 50~GeV, respectively, because the optimal FF pulse lengths span fewer multiples of the increased Rayleigh length $\eta_0$ than in the tighter-focusing case (see also Fig. \ref{Comparison1&5J}). As noted above, FF-enhanced pair production benefits strongly from tight focusing because it allows the pulse length to become much longer than $\eta_0$.

In conclusion, we have shown that a Joule-class FF pulse can significantly enhance the pair-production yield compared to a SFG pulse of the same energy. The enhancement originates from the more favorable scaling of the NBWPP probability with pulse duration than with laser intensity in the parameter regime considered. To maximize this advantage, high-energy photons with energies in the range $\sim 20$-50~GeV are required. Such photons can be produced via ICS using electron beams of comparable energies \cite{E144:1996enr, Burke:1997ew,LEP,XFEL2007,Shaw2025OPAL100GeV}. Our results indicate that the enhancement ratio increases with photon energy, suggesting that the enhancement may remain significant even for multi-Joule pulses using photons with $\sim 100$~GeV energy. We find that, for $\omega=50~$GeV and a 1-J FF pulse, the field intensity required to achieve the maximum pair-production yield is of order $10^{20}~\text{W/cm}^2$, which is one order of magnitude above the maximum intensity of a FF pulse produced to date \cite{Liberman2026PRR, Liberman2026FFLWFA}. Joule-class lasers are typically more widely available and affordable than higher-energy systems. They can also operate at kHz repetition rates, making them appealing for being combined with large-scale accelerators \cite{Palmer:25, KBELLA2017, Kiani2023}.

\acknowledgments

A.D.P. is partially supported by the U.S. National Science Foundation Mid-scale Research Infrastructure Program under Award No. PHY-2329970.

This material is based upon work supported by the U.S. Department of Energy [National Nuclear Security Administration] University of Rochester ``National Inertial Confinement Fusion Program'' under Award Number DE-NA0004144.

This report was prepared as an account of work sponsored by an agency of the United States Government. Neither the United States Government nor any agency thereof, nor any of their employees, makes any warranty, express or implied, or assumes any legal liability or responsibility for the accuracy, completeness, or usefulness of any information, apparatus, product, or process disclosed, or represents that its use would not infringe privately owned rights. Reference herein to any specific commercial product, process, or service by trade name, trademark, manufacturer, or otherwise does not necessarily constitute or imply its endorsement, recommendation, or favoring by the United States Government or any agency thereof. The views and opinions of authors expressed herein do not necessarily state or reflect those of the United States Government or any agency thereof.

\appendix*

\onecolumngrid

\section*{SUPPLEMENTAL MATERIAL}

\section*{Ptarmigan Implementation} \label{Fields}
The exact fields of the FF beam, linearly polarized along the $x$ axis with the central wave vector pointing along the positive $z$ axis and with focal velocity equal to the speed of light directed along the negative $z$ axis, are given by \cite{Formanek:2021bpw}
\begin{align}
        E_x(x,y,z,t) &= E_0 e^{-r^2/\sigma_\eta^2} (T_1+T_2),
\label{FFeqn1}\\ 
        E_y(x,y,z,t) &=\frac{2E_0xy}{\omega_0^2\sigma_0\sigma_\eta^3}e^{-r^2/\sigma_\eta^2}\sin(\Psi_2),\\
        E_z(x,y,z,t) &= \frac{2 E_0 x}{\omega_0\sigma_\eta^2}e^{-r^2/\sigma_\eta^2}\cos(\Psi_1), \label{EfieldZ}\\
        B_x(x,y,z,t) &= E_y(x,y,z,t)=E_y(y,x,z,t),\\
        B_y(x,y,z,t) &= E_x(y,x,z,t), \\
        B_z(x,y,z,t) &= E_z(y,x,z,t), \label{FFeqnEnd}
\end{align}
where
\begin{equation}
    T_1=\frac{\sigma_0}{\sigma_\eta} \sin(\Psi_0),\quad T_2 = \frac{x^2-y^2}{\omega_0^2 \sigma_0 \sigma_\eta^3}\sin\left(\Psi_2\right).
\label{FFeqn2}
\end{equation}
In Eqs.~(\ref{FFeqn1})--(\ref{FFeqn2}), we employ light-cone coordinates $\phi=t-z$ and $\eta=t+z$, together with $x$ and $y$ ($r=\sqrt{x^2+y^2}$), and the phases are
\begin{equation} \label{phase}
    \Psi_a = \omega_0 \phi - \frac{r^2}{\sigma_\eta^2} \frac{\eta}{\eta_0} + \big(1 + a \big) \arctan \bigg(\frac{\eta}{\eta_0} \bigg),
\end{equation}
where the integer $a$ determines the coefficient of the Gouy phase. The spot radius is $\sigma_\eta=\sigma_0\sqrt{1+\eta^2/\eta_0^2}$, and the Rayleigh length is $\eta_0=\omega_0\sigma_0^2$, with $\sigma_0$ being the laser waist. Importantly, the FF beam above is an exact solution of Maxwell's equations in vacuum.

The electromagnetic fields of the SFG pulse within the paraxial approximation are well known. We do not reproduce them here but we refer the reader to Ref.~\cite{Salamin_2007}, where expressions up to 10th order in the paraxial expansion are reported. We recall that the paraxial approximation relies on the smallness of the parameter $\sigma_0/z_0$ (hereafter referred to as the diffraction angle), where $z_0=\omega_0\sigma_0^2/2$ is the Rayleigh length of the SFG beam.

To obtain a pulsed FF beam, we multiply Eqs.~(\ref{FFeqn1})--(\ref{FFeqnEnd}) by the ramped flattop envelope defined in the Supplemental Material of Ref.~\cite{Formanek:2021bpw}. This provides the leading-order correction for finite-energy fields \cite{ramsey2023exact}. The SFG pulse is implemented in Ptarmigan to fourth order in the paraxial approximation. For consistency in the comparison with the FF pulse, we changed the default SFG ramp to the envelope used above. All pulses considered here have the same temporal profile.

Ptarmigan chooses the timestep automatically to maintain the accuracy of the Monte Carlo simulations \cite{Blackburn2023}.

At leading order in the paraxial approximation, the quantity $\mathcal{P}_{\text{avg}}=\frac{\pi}{4}(m/e)^2\xi_0^2\sigma_0^2\omega_0^2$ gives the average power of both SFG and FF pulses. For the tight focusing with $\sigma_0=1.01\lambda_0$ used in the simulations, the corresponding $\xi_0$ is overestimated by only $1.30\%$ for the SFG pulse \cite{Salamin_2007}. For the FF pulse, $\xi_0$ is underestimated by only $0.015\%$ for the same focusing \cite{Formanek:2021bpw}. Although both errors are small, the difference arises because the next-to-leading-order correction to the average power is second order in the diffraction angle for the SFG pulse, whereas it is fourth order for the FF pulse.

\section*{Benchmark of the simulations}
\label{Benchmark}

\begin{figure}[]
	\begin{center}
	\includegraphics[width=0.5\linewidth]{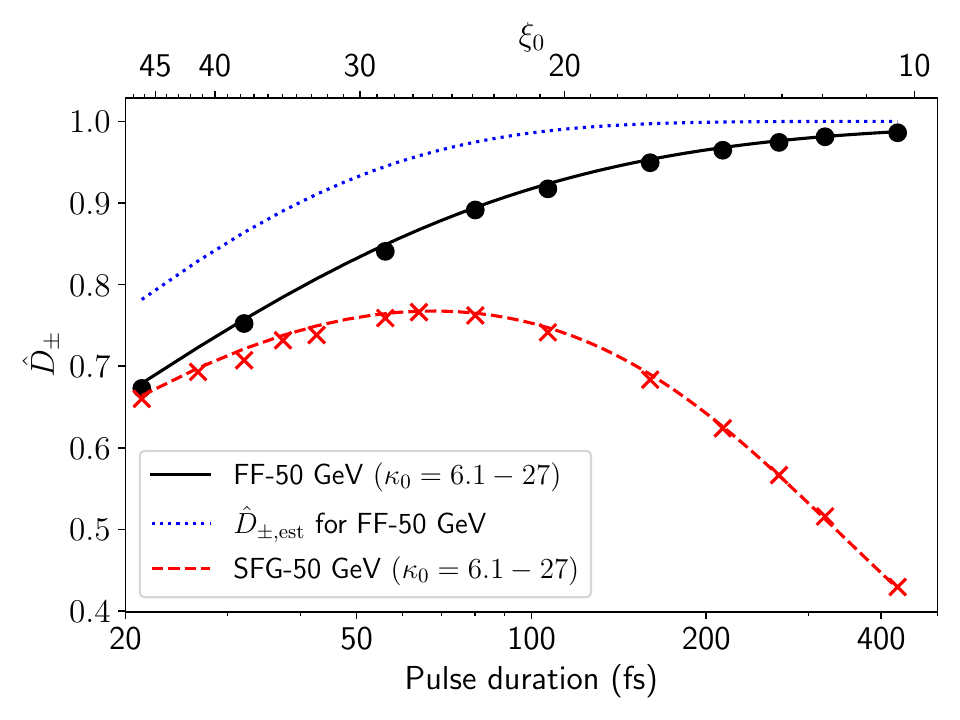}
	\end{center}
        \vspace{-0.8cm}
	\caption{Decay probability of the primary photons, $\hat{D}_\pm$, for FF and SFG pulses with the same pulse parameters as in Fig.~2, for $\omega=50$~GeV. The solid and dashed curves are obtained by numerically integrating Eq.~(1). The dotted curve shows the analytical estimate from Eq.~(2).} 
    \label{Decays}
\end{figure}

The decay probability of a primary photon can be estimated as $\hat{D}_{\pm,\text{est}}\approx 1-e^{-P}$, where $P$ is the time integral of the interpolated rate in Eq.~(1). We compare the numerical integrals of Eq.~(1) for the FF and SFG pulses at $\omega=50~$GeV with the simulation results in Fig.~\ref{Decays}.
The simulations show excellent agreement with the numerical integrals of Eq.~(1) at the percent level. In addition, Fig.~\ref{Decays} shows the analytical estimate $\hat{D}_{\pm,\text{est}}$ obtained from Eq.~(2) for the FF pulse. The deviation from the numerical integral arises because the $\kappa\gg 1$ asymptotic of Eq.~(1) overestimates the rate by approximately $\sim 25$--$100\%$ as $\kappa_0$ decreases from $\sim 27$ down to 6.1. For longer pulses this difference is less visible because both curves approach unity. Nevertheless, for the shortest pulses and largest $\kappa_0$, $\hat{D}_{\pm,\text{est}}$ still captures the correct scaling of the decay probability.

\section*{Energy and generation number of decayed photons for 1-J and 5-J pulses using 50-GeV photons} \label{decayed-photons-dist}

\begin{figure}
    \centering
    \includegraphics[width=0.49\linewidth]{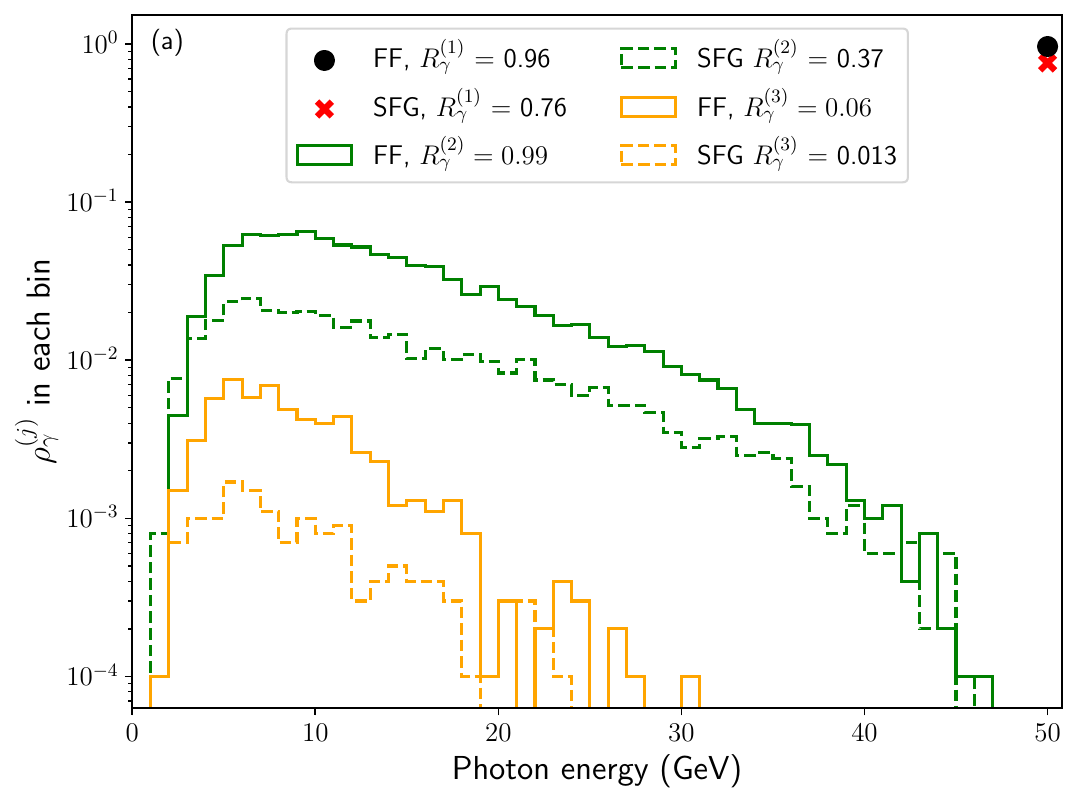}
    \hfill
    \includegraphics[width=0.49\linewidth]{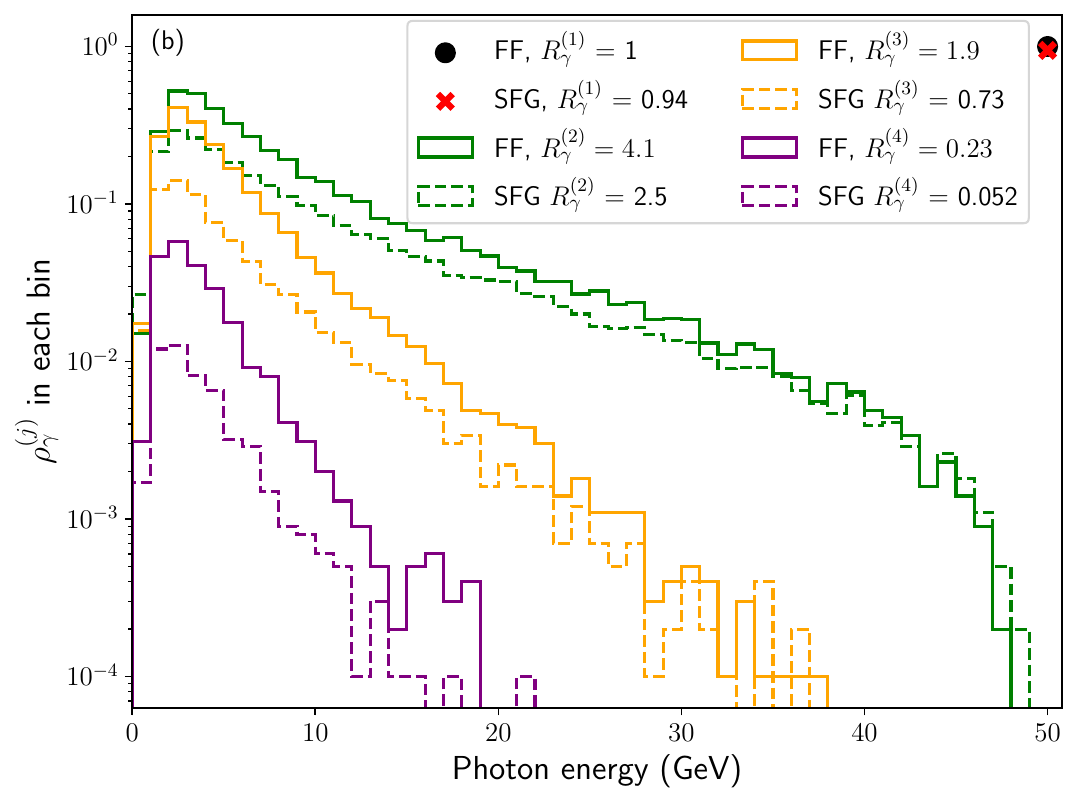}
    \caption{Energy distribution of $j$th-generation decayed photons per incoming photon for (a) 1~J and (b) 5~J FF and SFG pulses at their respective optimal parameters, using primary photons with $\omega=50~$GeV. The photon-energy bin size is 1~GeV. The quantities $R_{\gamma}^{(j)}$ in the legend indicate the total fraction of photons in the $j$th generation that decayed, i.e., the sum of $\rho_{\gamma}^{(j)}$ over all bins.}
    \label{photons1J}
\end{figure}

The energy distribution of decayed photons per incoming photon, $\rho_{\gamma}^{(j)}$, for generations $j=1,2,\ldots$ is shown in Fig.~\ref{photons1J}(a) for the 1-J pulse and in Fig.~\ref{photons1J}(b) for the 5-J pulse. The black dot and red cross indicate the primary decayed photons, $\rho_{\gamma}^{(1)}$ (first generation), all of which have 50-GeV energy, in the case of the FF pulse and the SFG pulse, respectively. The green curves show $\rho_{\gamma}^{(2)}$ in each photon-energy bin, corresponding to the second-generation decayed photons, i.e., photons emitted by electrons and positrons produced by first-generation photons, and so on.
We use the pulse parameters that yield the maximum electron--positron pair yield for these energies. The FF pulse produces slightly more pairs than the SFG pulse at every generation, leading to a cumulative enhancement. Figure~\ref{photons1J} also shows that the number of pairs per incoming photon in each bin decreases with increasing generation number. For 1~J pulses we show up to three generations, whereas for 5~J pulses we show up to four generations.

\twocolumngrid

%


\begin{thebibliography}{60}%
\makeatletter
\providecommand \@ifxundefined [1]{%
 \@ifx{#1\undefined}
}%
\providecommand \@ifnum [1]{%
 \ifnum #1\expandafter \@firstoftwo
 \else \expandafter \@secondoftwo
 \fi
}%
\providecommand \@ifx [1]{%
 \ifx #1\expandafter \@firstoftwo
 \else \expandafter \@secondoftwo
 \fi
}%
\providecommand \natexlab [1]{#1}%
\providecommand \enquote  [1]{``#1''}%
\providecommand \bibnamefont  [1]{#1}%
\providecommand \bibfnamefont [1]{#1}%
\providecommand \citenamefont [1]{#1}%
\providecommand \href@noop [0]{\@secondoftwo}%
\providecommand \href [0]{\begingroup \@sanitize@url \@href}%
\providecommand \@href[1]{\@@startlink{#1}\@@href}%
\providecommand \@@href[1]{\endgroup#1\@@endlink}%
\providecommand \@sanitize@url [0]{\catcode `\\12\catcode `\$12\catcode
  `\&12\catcode `\#12\catcode `\^12\catcode `\_12\catcode `\%12\relax}%
\providecommand \@@startlink[1]{}%
\providecommand \@@endlink[0]{}%
\providecommand \url  [0]{\begingroup\@sanitize@url \@url }%
\providecommand \@url [1]{\endgroup\@href {#1}{\urlprefix }}%
\providecommand \urlprefix  [0]{URL }%
\providecommand \Eprint [0]{\href }%
\providecommand \doibase [0]{https://doi.org/}%
\providecommand \selectlanguage [0]{\@gobble}%
\providecommand \bibinfo  [0]{\@secondoftwo}%
\providecommand \bibfield  [0]{\@secondoftwo}%
\providecommand \translation [1]{[#1]}%
\providecommand \BibitemOpen [0]{}%
\providecommand \bibitemStop [0]{}%
\providecommand \bibitemNoStop [0]{.\EOS\space}%
\providecommand \EOS [0]{\spacefactor3000\relax}%
\providecommand \BibitemShut  [1]{\csname bibitem#1\endcsname}%
\let\auto@bib@innerbib\@empty
\bibitem [{\citenamefont {Ritus}(1985)}]{Ritus:1985vta}%
  \BibitemOpen
  \bibfield  {author} {\bibinfo {author} {\bibfnamefont {V.~I.}\ \bibnamefont
  {Ritus}},\ }\href {https://doi.org/10.1007/BF01120220} {\bibfield  {journal}
  {\bibinfo  {journal} {J. Sov. Laser Res.}\ }\textbf {\bibinfo {volume} {6}},\
  \bibinfo {pages} {497} (\bibinfo {year} {1985})}\BibitemShut {NoStop}%
\bibitem [{\citenamefont {Baier}\ \emph {et~al.}(1998)\citenamefont {Baier},
  \citenamefont {Katkov},\ and\ \citenamefont {Strakhovenko}}]{Baier_b_1998}%
  \BibitemOpen
  \bibfield  {author} {\bibinfo {author} {\bibfnamefont {V.~N.}\ \bibnamefont
  {Baier}}, \bibinfo {author} {\bibfnamefont {V.~M.}\ \bibnamefont {Katkov}},\
  and\ \bibinfo {author} {\bibfnamefont {V.~M.}\ \bibnamefont {Strakhovenko}},\
  }\href@noop {} {\emph {\bibinfo {title} {Electromagnetic processes at high
  energies in oriented single crystals}}}\ (\bibinfo  {publisher} {World
  Scientific, Singapore},\ \bibinfo {year} {1998})\BibitemShut {NoStop}%
\bibitem [{\citenamefont {Di~Piazza}\ \emph {et~al.}(2012)\citenamefont
  {Di~Piazza}, \citenamefont {M{\" u}ller}, \citenamefont {Hatsagortsyan},\
  and\ \citenamefont {Keitel}}]{DiPiazza:2011tq}%
  \BibitemOpen
  \bibfield  {author} {\bibinfo {author} {\bibfnamefont {A.}~\bibnamefont
  {Di~Piazza}}, \bibinfo {author} {\bibfnamefont {C.}~\bibnamefont {M{\"
  u}ller}}, \bibinfo {author} {\bibfnamefont {K.~Z.}\ \bibnamefont
  {Hatsagortsyan}},\ and\ \bibinfo {author} {\bibfnamefont {C.~H.}\
  \bibnamefont {Keitel}},\ }\href {https://doi.org/10.1103/RevModPhys.84.1177}
  {\bibfield  {journal} {\bibinfo  {journal} {Rev. Mod. Phys.}\ }\textbf
  {\bibinfo {volume} {84}},\ \bibinfo {pages} {1177} (\bibinfo {year}
  {2012})}\BibitemShut {NoStop}%
\bibitem [{\citenamefont {Gonoskov}\ \emph {et~al.}(2022)\citenamefont
  {Gonoskov}, \citenamefont {Blackburn}, \citenamefont {Marklund},\ and\
  \citenamefont {Bulanov}}]{Gonoskov:2021hwf}%
  \BibitemOpen
  \bibfield  {author} {\bibinfo {author} {\bibfnamefont {A.}~\bibnamefont
  {Gonoskov}}, \bibinfo {author} {\bibfnamefont {T.~G.}\ \bibnamefont
  {Blackburn}}, \bibinfo {author} {\bibfnamefont {M.}~\bibnamefont
  {Marklund}},\ and\ \bibinfo {author} {\bibfnamefont {S.~S.}\ \bibnamefont
  {Bulanov}},\ }\href {https://doi.org/10.1103/RevModPhys.94.045001} {\bibfield
   {journal} {\bibinfo  {journal} {Rev. Mod. Phys.}\ }\textbf {\bibinfo
  {volume} {94}},\ \bibinfo {pages} {045001} (\bibinfo {year}
  {2022})}\BibitemShut {NoStop}%
\bibitem [{\citenamefont {Fedotov}\ \emph {et~al.}(2023)\citenamefont
  {Fedotov}, \citenamefont {Ilderton}, \citenamefont {Karbstein}, \citenamefont
  {King}, \citenamefont {Seipt}, \citenamefont {Taya},\ and\ \citenamefont
  {Torgrimsson}}]{Fedotov:2022ely}%
  \BibitemOpen
  \bibfield  {author} {\bibinfo {author} {\bibfnamefont {A.}~\bibnamefont
  {Fedotov}}, \bibinfo {author} {\bibfnamefont {A.}~\bibnamefont {Ilderton}},
  \bibinfo {author} {\bibfnamefont {F.}~\bibnamefont {Karbstein}}, \bibinfo
  {author} {\bibfnamefont {B.}~\bibnamefont {King}}, \bibinfo {author}
  {\bibfnamefont {D.}~\bibnamefont {Seipt}}, \bibinfo {author} {\bibfnamefont
  {H.}~\bibnamefont {Taya}},\ and\ \bibinfo {author} {\bibfnamefont
  {G.}~\bibnamefont {Torgrimsson}},\ }\href
  {https://doi.org/10.1016/j.physrep.2023.01.003} {\bibfield  {journal}
  {\bibinfo  {journal} {Phys. Rep.}\ }\textbf {\bibinfo {volume} {1010}},\
  \bibinfo {pages} {1} (\bibinfo {year} {2023})}\BibitemShut {NoStop}%
\bibitem [{\citenamefont {Strickland}\ and\ \citenamefont
  {Mourou}(1985)}]{Strickland_1985}%
  \BibitemOpen
  \bibfield  {author} {\bibinfo {author} {\bibfnamefont {D.}~\bibnamefont
  {Strickland}}\ and\ \bibinfo {author} {\bibfnamefont {G.}~\bibnamefont
  {Mourou}},\ }\href@noop {} {\bibfield  {journal} {\bibinfo  {journal} {Optics
  Commun.}\ }\textbf {\bibinfo {volume} {56}},\ \bibinfo {pages} {219}
  (\bibinfo {year} {1985})}\BibitemShut {NoStop}%
\bibitem [{\citenamefont {Danson}\ \emph {et~al.}(2019)\citenamefont {Danson},
  \citenamefont {Haefner}, \citenamefont {Bromage}, \citenamefont {Butcher},
  \citenamefont {Chanteloup}, \citenamefont {Chowdhury}, \citenamefont
  {Galvanauskas}, \citenamefont {Gizzi}, \citenamefont {Hein}, \citenamefont
  {Hillier} \emph {et~al.}}]{Danson2019}%
  \BibitemOpen
  \bibfield  {author} {\bibinfo {author} {\bibfnamefont {C.~N.}\ \bibnamefont
  {Danson}}, \bibinfo {author} {\bibfnamefont {C.}~\bibnamefont {Haefner}},
  \bibinfo {author} {\bibfnamefont {J.}~\bibnamefont {Bromage}}, \bibinfo
  {author} {\bibfnamefont {T.}~\bibnamefont {Butcher}}, \bibinfo {author}
  {\bibfnamefont {J.-C.~F.}\ \bibnamefont {Chanteloup}}, \bibinfo {author}
  {\bibfnamefont {E.~A.}\ \bibnamefont {Chowdhury}}, \bibinfo {author}
  {\bibfnamefont {A.}~\bibnamefont {Galvanauskas}}, \bibinfo {author}
  {\bibfnamefont {L.~A.}\ \bibnamefont {Gizzi}}, \bibinfo {author}
  {\bibfnamefont {J.}~\bibnamefont {Hein}}, \bibinfo {author} {\bibfnamefont
  {D.~I.}\ \bibnamefont {Hillier}}, \emph {et~al.},\ }\href
  {https://www.researching.cn/articles/OJ2151e170fd94be69} {\bibfield
  {journal} {\bibinfo  {journal} {High Power Laser Science and Engineering}\
  }\textbf {\bibinfo {volume} {7}},\ \bibinfo {pages} {03000e54} (\bibinfo
  {year} {2019})}\BibitemShut {NoStop}%
\bibitem [{\citenamefont {Yoon}\ \emph {et~al.}(2021)\citenamefont {Yoon},
  \citenamefont {Kim}, \citenamefont {Choi}, \citenamefont {Sung},
  \citenamefont {Lee}, \citenamefont {Lee},\ and\ \citenamefont
  {Nam}}]{Yoon:2021ony}%
  \BibitemOpen
  \bibfield  {author} {\bibinfo {author} {\bibfnamefont {J.~W.}\ \bibnamefont
  {Yoon}}, \bibinfo {author} {\bibfnamefont {Y.~G.}\ \bibnamefont {Kim}},
  \bibinfo {author} {\bibfnamefont {I.~W.}\ \bibnamefont {Choi}}, \bibinfo
  {author} {\bibfnamefont {J.~H.}\ \bibnamefont {Sung}}, \bibinfo {author}
  {\bibfnamefont {H.~W.}\ \bibnamefont {Lee}}, \bibinfo {author} {\bibfnamefont
  {S.~K.}\ \bibnamefont {Lee}},\ and\ \bibinfo {author} {\bibfnamefont {C.~H.}\
  \bibnamefont {Nam}},\ }\href {https://doi.org/10.1364/OPTICA.420520}
  {\bibfield  {journal} {\bibinfo  {journal} {Optica}\ }\textbf {\bibinfo
  {volume} {8}},\ \bibinfo {pages} {630} (\bibinfo {year} {2021})}\BibitemShut
  {NoStop}%
\bibitem [{\citenamefont {Bula}\ \emph {et~al.}(1996)\citenamefont {Bula} \emph
  {et~al.}}]{E144:1996enr}%
  \BibitemOpen
  \bibfield  {author} {\bibinfo {author} {\bibfnamefont {C.}~\bibnamefont
  {Bula}} \emph {et~al.} (\bibinfo {collaboration} {E144}),\ }\href
  {https://doi.org/10.1103/PhysRevLett.76.3116} {\bibfield  {journal} {\bibinfo
   {journal} {Phys. Rev. Lett.}\ }\textbf {\bibinfo {volume} {76}},\ \bibinfo
  {pages} {3116} (\bibinfo {year} {1996})}\BibitemShut {NoStop}%
\bibitem [{\citenamefont {Cole}\ \emph {et~al.}(2018)\citenamefont {Cole},
  \citenamefont {Behm}, \citenamefont {Gerstmayr}, \citenamefont {Blackburn},
  \citenamefont {Wood}, \citenamefont {Baird}, \citenamefont {Duff},
  \citenamefont {Harvey}, \citenamefont {Ilderton}, \citenamefont {Joglekar}
  \emph {et~al.}}]{Cole_2018}%
  \BibitemOpen
  \bibfield  {author} {\bibinfo {author} {\bibfnamefont {J.~M.}\ \bibnamefont
  {Cole}}, \bibinfo {author} {\bibfnamefont {K.~T.}\ \bibnamefont {Behm}},
  \bibinfo {author} {\bibfnamefont {E.}~\bibnamefont {Gerstmayr}}, \bibinfo
  {author} {\bibfnamefont {T.~G.}\ \bibnamefont {Blackburn}}, \bibinfo {author}
  {\bibfnamefont {J.~C.}\ \bibnamefont {Wood}}, \bibinfo {author}
  {\bibfnamefont {C.~D.}\ \bibnamefont {Baird}}, \bibinfo {author}
  {\bibfnamefont {M.~J.}\ \bibnamefont {Duff}}, \bibinfo {author}
  {\bibfnamefont {C.}~\bibnamefont {Harvey}}, \bibinfo {author} {\bibfnamefont
  {A.}~\bibnamefont {Ilderton}}, \bibinfo {author} {\bibfnamefont {A.~S.}\
  \bibnamefont {Joglekar}}, \emph {et~al.},\ }\href
  {https://doi.org/10.1103/PhysRevX.8.011020} {\bibfield  {journal} {\bibinfo
  {journal} {Phys. Rev. X}\ }\textbf {\bibinfo {volume} {8}},\ \bibinfo {pages}
  {011020} (\bibinfo {year} {2018})}\BibitemShut {NoStop}%
\bibitem [{\citenamefont {Poder}\ \emph {et~al.}(2018)\citenamefont {Poder},
  \citenamefont {Tamburini}, \citenamefont {Sarri}, \citenamefont {Di~Piazza},
  \citenamefont {Kuschel}, \citenamefont {Baird}, \citenamefont {Behm},
  \citenamefont {Bohlen}, \citenamefont {Cole}, \citenamefont {Corvan} \emph
  {et~al.}}]{Poder_2018}%
  \BibitemOpen
  \bibfield  {author} {\bibinfo {author} {\bibfnamefont {K.}~\bibnamefont
  {Poder}}, \bibinfo {author} {\bibfnamefont {M.}~\bibnamefont {Tamburini}},
  \bibinfo {author} {\bibfnamefont {G.}~\bibnamefont {Sarri}}, \bibinfo
  {author} {\bibfnamefont {A.}~\bibnamefont {Di~Piazza}}, \bibinfo {author}
  {\bibfnamefont {S.}~\bibnamefont {Kuschel}}, \bibinfo {author} {\bibfnamefont
  {C.~D.}\ \bibnamefont {Baird}}, \bibinfo {author} {\bibfnamefont
  {K.}~\bibnamefont {Behm}}, \bibinfo {author} {\bibfnamefont {S.}~\bibnamefont
  {Bohlen}}, \bibinfo {author} {\bibfnamefont {J.~M.}\ \bibnamefont {Cole}},
  \bibinfo {author} {\bibfnamefont {D.~J.}\ \bibnamefont {Corvan}}, \emph
  {et~al.},\ }\href {https://doi.org/10.1103/PhysRevX.8.031004} {\bibfield
  {journal} {\bibinfo  {journal} {Phys. Rev. X}\ }\textbf {\bibinfo {volume}
  {8}},\ \bibinfo {pages} {031004} (\bibinfo {year} {2018})}\BibitemShut
  {NoStop}%
\bibitem [{\citenamefont {Mirzaie}\ \emph {et~al.}(2024)\citenamefont
  {Mirzaie}, \citenamefont {Hojbota}, \citenamefont {Kim}, \citenamefont
  {Pathak}, \citenamefont {Pak}, \citenamefont {Kim}, \citenamefont {Lee},
  \citenamefont {Yoon}, \citenamefont {Lee}, \citenamefont {Rhee} \emph
  {et~al.}}]{Mirzaie_2024}%
  \BibitemOpen
  \bibfield  {author} {\bibinfo {author} {\bibfnamefont {M.}~\bibnamefont
  {Mirzaie}}, \bibinfo {author} {\bibfnamefont {C.~I.}\ \bibnamefont
  {Hojbota}}, \bibinfo {author} {\bibfnamefont {D.~Y.}\ \bibnamefont {Kim}},
  \bibinfo {author} {\bibfnamefont {V.~B.}\ \bibnamefont {Pathak}}, \bibinfo
  {author} {\bibfnamefont {T.~G.}\ \bibnamefont {Pak}}, \bibinfo {author}
  {\bibfnamefont {C.~M.}\ \bibnamefont {Kim}}, \bibinfo {author} {\bibfnamefont
  {H.~W.}\ \bibnamefont {Lee}}, \bibinfo {author} {\bibfnamefont {J.~W.}\
  \bibnamefont {Yoon}}, \bibinfo {author} {\bibfnamefont {S.~K.}\ \bibnamefont
  {Lee}}, \bibinfo {author} {\bibfnamefont {Y.~J.}\ \bibnamefont {Rhee}}, \emph
  {et~al.},\ }\href {https://doi.org/10.1038/s41566-024-01550-8} {\bibfield
  {journal} {\bibinfo  {journal} {Nature Photon.}\ }\textbf {\bibinfo {volume}
  {18}},\ \bibinfo {pages} {1212} (\bibinfo {year} {2024})}\BibitemShut
  {NoStop}%
\bibitem [{\citenamefont {Los}\ \emph {et~al.}(2026)\citenamefont {Los},
  \citenamefont {Gerstmayr}, \citenamefont {Arran}, \citenamefont {Streeter},
  \citenamefont {Colgan}, \citenamefont {Cobo}, \citenamefont {Kettle},
  \citenamefont {Blackburn}, \citenamefont {Bourgeois}, \citenamefont {Calvin}
  \emph {et~al.}}]{Los_2026}%
  \BibitemOpen
  \bibfield  {author} {\bibinfo {author} {\bibfnamefont {E.~E.}\ \bibnamefont
  {Los}}, \bibinfo {author} {\bibfnamefont {E.}~\bibnamefont {Gerstmayr}},
  \bibinfo {author} {\bibfnamefont {C.}~\bibnamefont {Arran}}, \bibinfo
  {author} {\bibfnamefont {M.~J.~V.}\ \bibnamefont {Streeter}}, \bibinfo
  {author} {\bibfnamefont {C.}~\bibnamefont {Colgan}}, \bibinfo {author}
  {\bibfnamefont {C.~C.}\ \bibnamefont {Cobo}}, \bibinfo {author}
  {\bibfnamefont {B.}~\bibnamefont {Kettle}}, \bibinfo {author} {\bibfnamefont
  {T.~G.}\ \bibnamefont {Blackburn}}, \bibinfo {author} {\bibfnamefont
  {N.}~\bibnamefont {Bourgeois}}, \bibinfo {author} {\bibfnamefont
  {L.}~\bibnamefont {Calvin}}, \emph {et~al.},\ }\href
  {https://doi.org/10.1038/s41467-025-67918-8} {\bibfield  {journal} {\bibinfo
  {journal} {Nature Commun.}\ }\textbf {\bibinfo {volume} {17}},\ \bibinfo
  {pages} {1157} (\bibinfo {year} {2026})}\BibitemShut {NoStop}%
\bibitem [{\citenamefont {Burke}\ \emph {et~al.}(1997)\citenamefont {Burke},
  \citenamefont {Field}, \citenamefont {Horton-Smith}, \citenamefont {Spencer},
  \citenamefont {Walz}, \citenamefont {Berridge}, \citenamefont {Bugg},
  \citenamefont {Shmakov}, \citenamefont {Weidemann}, \citenamefont {C.} \emph
  {et~al.}}]{Burke:1997ew}%
  \BibitemOpen
  \bibfield  {author} {\bibinfo {author} {\bibfnamefont {D.~L.}\ \bibnamefont
  {Burke}}, \bibinfo {author} {\bibfnamefont {R.~C.}\ \bibnamefont {Field}},
  \bibinfo {author} {\bibfnamefont {G.}~\bibnamefont {Horton-Smith}}, \bibinfo
  {author} {\bibfnamefont {J.~E.}\ \bibnamefont {Spencer}}, \bibinfo {author}
  {\bibfnamefont {D.}~\bibnamefont {Walz}}, \bibinfo {author} {\bibfnamefont
  {S.~C.}\ \bibnamefont {Berridge}}, \bibinfo {author} {\bibfnamefont {W.~M.}\
  \bibnamefont {Bugg}}, \bibinfo {author} {\bibfnamefont {K.}~\bibnamefont
  {Shmakov}}, \bibinfo {author} {\bibfnamefont {A.~W.}\ \bibnamefont
  {Weidemann}}, \bibinfo {author} {\bibfnamefont {B.}~\bibnamefont {C.}}, \emph
  {et~al.},\ }\href {https://doi.org/10.1103/PhysRevLett.79.1626} {\bibfield
  {journal} {\bibinfo  {journal} {Phys. Rev. Lett.}\ }\textbf {\bibinfo
  {volume} {79}},\ \bibinfo {pages} {1626} (\bibinfo {year}
  {1997})}\BibitemShut {NoStop}%
\bibitem [{\citenamefont {Weber}\ \emph {et~al.}(2017)\citenamefont {Weber},
  \citenamefont {Bechet}, \citenamefont {Borneis}, \citenamefont {Brabec},
  \citenamefont {Bucka}, \citenamefont {Chacon-Golcher}, \citenamefont
  {Ciappina}, \citenamefont {DeMarco}, \citenamefont {Fajstavr}, \citenamefont
  {Falk} \emph {et~al.}}]{Weber2017}%
  \BibitemOpen
  \bibfield  {author} {\bibinfo {author} {\bibfnamefont {S.}~\bibnamefont
  {Weber}}, \bibinfo {author} {\bibfnamefont {S.}~\bibnamefont {Bechet}},
  \bibinfo {author} {\bibfnamefont {S.}~\bibnamefont {Borneis}}, \bibinfo
  {author} {\bibfnamefont {L.}~\bibnamefont {Brabec}}, \bibinfo {author}
  {\bibfnamefont {M.}~\bibnamefont {Bucka}}, \bibinfo {author} {\bibfnamefont
  {E.}~\bibnamefont {Chacon-Golcher}}, \bibinfo {author} {\bibfnamefont
  {M.}~\bibnamefont {Ciappina}}, \bibinfo {author} {\bibfnamefont
  {M.}~\bibnamefont {DeMarco}}, \bibinfo {author} {\bibfnamefont
  {A.}~\bibnamefont {Fajstavr}}, \bibinfo {author} {\bibfnamefont
  {K.}~\bibnamefont {Falk}}, \emph {et~al.},\ }\href
  {https://doi.org/10.1016/j.mre.2017.03.003} {\bibfield  {journal} {\bibinfo
  {journal} {Matter Radiat. Extremes}\ }\textbf {\bibinfo {volume} {2}},\
  \bibinfo {pages} {149} (\bibinfo {year} {2017})}\BibitemShut {NoStop}%
\bibitem [{\citenamefont {Gales}\ \emph {et~al.}(2018)\citenamefont {Gales},
  \citenamefont {Tanaka}, \citenamefont {Balabanski}, \citenamefont {Negoita},
  \citenamefont {Stutman}, \citenamefont {Tesileanu}, \citenamefont {Ur},
  \citenamefont {Ursescu}, \citenamefont {Andrei}, \citenamefont {Ataman} \emph
  {et~al.}}]{Gales2018}%
  \BibitemOpen
  \bibfield  {author} {\bibinfo {author} {\bibfnamefont {S.}~\bibnamefont
  {Gales}}, \bibinfo {author} {\bibfnamefont {K.~A.}\ \bibnamefont {Tanaka}},
  \bibinfo {author} {\bibfnamefont {D.~L.}\ \bibnamefont {Balabanski}},
  \bibinfo {author} {\bibfnamefont {F.}~\bibnamefont {Negoita}}, \bibinfo
  {author} {\bibfnamefont {D.}~\bibnamefont {Stutman}}, \bibinfo {author}
  {\bibfnamefont {O.}~\bibnamefont {Tesileanu}}, \bibinfo {author}
  {\bibfnamefont {C.~A.}\ \bibnamefont {Ur}}, \bibinfo {author} {\bibfnamefont
  {D.}~\bibnamefont {Ursescu}}, \bibinfo {author} {\bibfnamefont
  {I.}~\bibnamefont {Andrei}}, \bibinfo {author} {\bibfnamefont
  {S.}~\bibnamefont {Ataman}}, \emph {et~al.},\ }\href
  {https://doi.org/10.1088/1361-6633/aacfe8} {\bibfield  {journal} {\bibinfo
  {journal} {Rep. Prog. Phys.}\ }\textbf {\bibinfo {volume} {81}},\ \bibinfo
  {pages} {094301} (\bibinfo {year} {2018})}\BibitemShut {NoStop}%
\bibitem [{\citenamefont {Nees}\ \emph {et~al.}(2020)\citenamefont {Nees},
  \citenamefont {Maksimchuk}, \citenamefont {Kalinchenko}, \citenamefont {Hou},
  \citenamefont {Ma}, \citenamefont {Campbell}, \citenamefont {McKelvey},
  \citenamefont {Willingale}, \citenamefont {Jovanovic}, \citenamefont {Kuranz}
  \emph {et~al.}}]{Nees2020}%
  \BibitemOpen
  \bibfield  {author} {\bibinfo {author} {\bibfnamefont {J.}~\bibnamefont
  {Nees}}, \bibinfo {author} {\bibfnamefont {A.}~\bibnamefont {Maksimchuk}},
  \bibinfo {author} {\bibfnamefont {G.}~\bibnamefont {Kalinchenko}}, \bibinfo
  {author} {\bibfnamefont {B.}~\bibnamefont {Hou}}, \bibinfo {author}
  {\bibfnamefont {Y.}~\bibnamefont {Ma}}, \bibinfo {author} {\bibfnamefont
  {P.}~\bibnamefont {Campbell}}, \bibinfo {author} {\bibfnamefont
  {A.}~\bibnamefont {McKelvey}}, \bibinfo {author} {\bibfnamefont
  {L.}~\bibnamefont {Willingale}}, \bibinfo {author} {\bibfnamefont
  {I.}~\bibnamefont {Jovanovic}}, \bibinfo {author} {\bibfnamefont
  {C.}~\bibnamefont {Kuranz}}, \emph {et~al.}\ }(\bibinfo  {publisher} {Optica
  Publishing Group},\ \bibinfo {year} {2020})\ p.\ \bibinfo {pages}
  {JW2B.9}\BibitemShut {NoStop}%
\bibitem [{NSF()}]{NSF_OPAL}%
  \BibitemOpen
  \href@noop {} {\bibinfo {title} {{National Science Foundation Optical
  Parametric Amplifier Lines (NSF OPAL)}}},\ \bibinfo {howpublished}
  {\url{https://nsf-opal.rochester.edu/}}\BibitemShut {NoStop}%
\bibitem [{\citenamefont {Vranic}\ \emph {et~al.}(2018)\citenamefont {Vranic},
  \citenamefont {Klimo}, \citenamefont {Korn},\ and\ \citenamefont
  {Weber}}]{Vranic2018}%
  \BibitemOpen
  \bibfield  {author} {\bibinfo {author} {\bibfnamefont {M.}~\bibnamefont
  {Vranic}}, \bibinfo {author} {\bibfnamefont {O.}~\bibnamefont {Klimo}},
  \bibinfo {author} {\bibfnamefont {G.}~\bibnamefont {Korn}},\ and\ \bibinfo
  {author} {\bibfnamefont {S.}~\bibnamefont {Weber}},\ }\href
  {https://doi.org/10.1038/s41598-018-23126-7} {\bibfield  {journal} {\bibinfo
  {journal} {Scientific Reports}\ }\textbf {\bibinfo {volume} {8}},\ \bibinfo
  {pages} {4702} (\bibinfo {year} {2018})}\BibitemShut {NoStop}%
\bibitem [{\citenamefont {He}\ \emph {et~al.}(2022)\citenamefont {He},
  \citenamefont {Blackburn}, \citenamefont {Toncian},\ and\ \citenamefont
  {Arefiev}}]{He2018}%
  \BibitemOpen
  \bibfield  {author} {\bibinfo {author} {\bibfnamefont {Y.}~\bibnamefont
  {He}}, \bibinfo {author} {\bibfnamefont {T.~G.}\ \bibnamefont {Blackburn}},
  \bibinfo {author} {\bibfnamefont {T.}~\bibnamefont {Toncian}},\ and\ \bibinfo
  {author} {\bibfnamefont {A.}~\bibnamefont {Arefiev}},\ }\href
  {https://doi.org/10.1063/5.0086577} {\bibfield  {journal} {\bibinfo
  {journal} {Physics of Plasmas}\ }\textbf {\bibinfo {volume} {29}},\ \bibinfo
  {pages} {053105} (\bibinfo {year} {2022})}\BibitemShut {NoStop}%
\bibitem [{\citenamefont {Sainte-Marie}\ \emph {et~al.}(2017)\citenamefont
  {Sainte-Marie}, \citenamefont {Gobert},\ and\ \citenamefont
  {Qu\'{e}r\'{e}}}]{Sainte-Marie_2017}%
  \BibitemOpen
  \bibfield  {author} {\bibinfo {author} {\bibfnamefont {A.}~\bibnamefont
  {Sainte-Marie}}, \bibinfo {author} {\bibfnamefont {O.}~\bibnamefont
  {Gobert}},\ and\ \bibinfo {author} {\bibfnamefont {F.}~\bibnamefont
  {Qu\'{e}r\'{e}}},\ }\href {https://doi.org/10.1364/OPTICA.4.001298}
  {\bibfield  {journal} {\bibinfo  {journal} {Optica}\ }\textbf {\bibinfo
  {volume} {4}},\ \bibinfo {pages} {1298} (\bibinfo {year} {2017})}\BibitemShut
  {NoStop}%
\bibitem [{\citenamefont {Froula}\ \emph {et~al.}(2018)\citenamefont {Froula},
  \citenamefont {Turnbull}, \citenamefont {Davies}, \citenamefont {Kessler},
  \citenamefont {Haberberger}, \citenamefont {Palastro}, \citenamefont {Bahk},
  \citenamefont {Begishev}, \citenamefont {Boni}, \citenamefont {Bucht} \emph
  {et~al.}}]{Froula_2018}%
  \BibitemOpen
  \bibfield  {author} {\bibinfo {author} {\bibfnamefont {D.~H.}\ \bibnamefont
  {Froula}}, \bibinfo {author} {\bibfnamefont {D.}~\bibnamefont {Turnbull}},
  \bibinfo {author} {\bibfnamefont {A.~S.}\ \bibnamefont {Davies}}, \bibinfo
  {author} {\bibfnamefont {T.~J.}\ \bibnamefont {Kessler}}, \bibinfo {author}
  {\bibfnamefont {D.}~\bibnamefont {Haberberger}}, \bibinfo {author}
  {\bibfnamefont {J.~P.}\ \bibnamefont {Palastro}}, \bibinfo {author}
  {\bibfnamefont {S.-W.}\ \bibnamefont {Bahk}}, \bibinfo {author}
  {\bibfnamefont {I.~A.}\ \bibnamefont {Begishev}}, \bibinfo {author}
  {\bibfnamefont {R.}~\bibnamefont {Boni}}, \bibinfo {author} {\bibfnamefont
  {S.}~\bibnamefont {Bucht}}, \emph {et~al.},\ }\href
  {https://doi.org/10.1038/s41566-018-0121-8} {\bibfield  {journal} {\bibinfo
  {journal} {Nat. Photonics}\ }\textbf {\bibinfo {volume} {12}},\ \bibinfo
  {pages} {262} (\bibinfo {year} {2018})}\BibitemShut {NoStop}%
\bibitem [{\citenamefont {Turnbull}\ \emph {et~al.}(2018)\citenamefont
  {Turnbull}, \citenamefont {Franke}, \citenamefont {Katz}, \citenamefont
  {Palastro}, \citenamefont {Begishev}, \citenamefont {Boni}, \citenamefont
  {Bromage}, \citenamefont {Milder}, \citenamefont {Shaw},\ and\ \citenamefont
  {Froula}}]{Turbull2018}%
  \BibitemOpen
  \bibfield  {author} {\bibinfo {author} {\bibfnamefont {D.}~\bibnamefont
  {Turnbull}}, \bibinfo {author} {\bibfnamefont {P.}~\bibnamefont {Franke}},
  \bibinfo {author} {\bibfnamefont {J.}~\bibnamefont {Katz}}, \bibinfo {author}
  {\bibfnamefont {J.~P.}\ \bibnamefont {Palastro}}, \bibinfo {author}
  {\bibfnamefont {I.~A.}\ \bibnamefont {Begishev}}, \bibinfo {author}
  {\bibfnamefont {R.}~\bibnamefont {Boni}}, \bibinfo {author} {\bibfnamefont
  {J.}~\bibnamefont {Bromage}}, \bibinfo {author} {\bibfnamefont {A.~L.}\
  \bibnamefont {Milder}}, \bibinfo {author} {\bibfnamefont {J.~L.}\
  \bibnamefont {Shaw}},\ and\ \bibinfo {author} {\bibfnamefont {D.~H.}\
  \bibnamefont {Froula}},\ }\href
  {https://doi.org/10.1103/PhysRevLett.120.225001} {\bibfield  {journal}
  {\bibinfo  {journal} {Phys. Rev. Lett.}\ }\textbf {\bibinfo {volume} {120}},\
  \bibinfo {pages} {225001} (\bibinfo {year} {2018})}\BibitemShut {NoStop}%
\bibitem [{\citenamefont {Jolly}\ \emph {et~al.}(2020)\citenamefont {Jolly},
  \citenamefont {Gobert}, \citenamefont {Jeandet},\ and\ \citenamefont
  {Qu\'{e}r\'{e}}}]{Jolly:2020}%
  \BibitemOpen
  \bibfield  {author} {\bibinfo {author} {\bibfnamefont {S.~W.}\ \bibnamefont
  {Jolly}}, \bibinfo {author} {\bibfnamefont {O.}~\bibnamefont {Gobert}},
  \bibinfo {author} {\bibfnamefont {A.}~\bibnamefont {Jeandet}},\ and\ \bibinfo
  {author} {\bibfnamefont {F.}~\bibnamefont {Qu\'{e}r\'{e}}},\ }\href
  {https://doi.org/10.1364/OE.384512} {\bibfield  {journal} {\bibinfo
  {journal} {Opt. Express}\ }\textbf {\bibinfo {volume} {28}},\ \bibinfo
  {pages} {4888} (\bibinfo {year} {2020})}\BibitemShut {NoStop}%
\bibitem [{\citenamefont {Simpson}\ \emph {et~al.}(2020)\citenamefont
  {Simpson}, \citenamefont {Ramsey}, \citenamefont {Franke}, \citenamefont
  {Vafaei-Najafabadi}, \citenamefont {Turnbull}, \citenamefont {Froula},\ and\
  \citenamefont {Palastro}}]{simpson2020nonlinear}%
  \BibitemOpen
  \bibfield  {author} {\bibinfo {author} {\bibfnamefont {T.~T.}\ \bibnamefont
  {Simpson}}, \bibinfo {author} {\bibfnamefont {D.}~\bibnamefont {Ramsey}},
  \bibinfo {author} {\bibfnamefont {P.}~\bibnamefont {Franke}}, \bibinfo
  {author} {\bibfnamefont {N.}~\bibnamefont {Vafaei-Najafabadi}}, \bibinfo
  {author} {\bibfnamefont {D.}~\bibnamefont {Turnbull}}, \bibinfo {author}
  {\bibfnamefont {D.~H.}\ \bibnamefont {Froula}},\ and\ \bibinfo {author}
  {\bibfnamefont {J.~P.}\ \bibnamefont {Palastro}},\ }\href
  {https://doi.org/10.1364/OE.411011} {\bibfield  {journal} {\bibinfo
  {journal} {Opt. Express}\ }\textbf {\bibinfo {volume} {28}},\ \bibinfo
  {pages} {38516} (\bibinfo {year} {2020})}\BibitemShut {NoStop}%
\bibitem [{\citenamefont {Simpson}\ \emph {et~al.}(2022)\citenamefont
  {Simpson}, \citenamefont {Ramsey}, \citenamefont {Franke}, \citenamefont
  {Weichman}, \citenamefont {Ambat}, \citenamefont {Turnbull}, \citenamefont
  {Froula},\ and\ \citenamefont {Palastro}}]{simpson2022spatiotemporal}%
  \BibitemOpen
  \bibfield  {author} {\bibinfo {author} {\bibfnamefont {T.~T.}\ \bibnamefont
  {Simpson}}, \bibinfo {author} {\bibfnamefont {D.}~\bibnamefont {Ramsey}},
  \bibinfo {author} {\bibfnamefont {P.}~\bibnamefont {Franke}}, \bibinfo
  {author} {\bibfnamefont {K.}~\bibnamefont {Weichman}}, \bibinfo {author}
  {\bibfnamefont {M.~V.}\ \bibnamefont {Ambat}}, \bibinfo {author}
  {\bibfnamefont {D.}~\bibnamefont {Turnbull}}, \bibinfo {author}
  {\bibfnamefont {D.~H.}\ \bibnamefont {Froula}},\ and\ \bibinfo {author}
  {\bibfnamefont {J.~P.}\ \bibnamefont {Palastro}},\ }\href
  {https://doi.org/10.1364/OE.451123} {\bibfield  {journal} {\bibinfo
  {journal} {Opt. Express}\ }\textbf {\bibinfo {volume} {30}},\ \bibinfo
  {pages} {9878} (\bibinfo {year} {2022})}\BibitemShut {NoStop}%
\bibitem [{\citenamefont {Palastro}\ \emph {et~al.}(2020)\citenamefont
  {Palastro}, \citenamefont {Shaw}, \citenamefont {Franke}, \citenamefont
  {Ramsey}, \citenamefont {Simpson},\ and\ \citenamefont
  {Froula}}]{Palastro:2020gcl}%
  \BibitemOpen
  \bibfield  {author} {\bibinfo {author} {\bibfnamefont {J.~P.}\ \bibnamefont
  {Palastro}}, \bibinfo {author} {\bibfnamefont {J.~L.}\ \bibnamefont {Shaw}},
  \bibinfo {author} {\bibfnamefont {P.}~\bibnamefont {Franke}}, \bibinfo
  {author} {\bibfnamefont {D.}~\bibnamefont {Ramsey}}, \bibinfo {author}
  {\bibfnamefont {T.~T.}\ \bibnamefont {Simpson}},\ and\ \bibinfo {author}
  {\bibfnamefont {D.~H.}\ \bibnamefont {Froula}},\ }\href
  {https://doi.org/10.1103/PhysRevLett.124.134802} {\bibfield  {journal}
  {\bibinfo  {journal} {Phys. Rev. Lett.}\ }\textbf {\bibinfo {volume} {124}},\
  \bibinfo {pages} {134802} (\bibinfo {year} {2020})},\ \bibinfo {note}
  {[Erratum: Phys.Rev.Lett. 130, 159902 (2023)]}\BibitemShut {NoStop}%
\bibitem [{\citenamefont {Pigeon}\ \emph {et~al.}(2024)\citenamefont {Pigeon},
  \citenamefont {Franke}, \citenamefont {Chong}, \citenamefont {Katz},
  \citenamefont {Boni}, \citenamefont {Dorrer}, \citenamefont {Palastro},\ and\
  \citenamefont {Froula}}]{pigeon2024ultrabroadband}%
  \BibitemOpen
  \bibfield  {author} {\bibinfo {author} {\bibfnamefont {J.}~\bibnamefont
  {Pigeon}}, \bibinfo {author} {\bibfnamefont {P.}~\bibnamefont {Franke}},
  \bibinfo {author} {\bibfnamefont {M.~L.~P.}\ \bibnamefont {Chong}}, \bibinfo
  {author} {\bibfnamefont {J.}~\bibnamefont {Katz}}, \bibinfo {author}
  {\bibfnamefont {R.}~\bibnamefont {Boni}}, \bibinfo {author} {\bibfnamefont
  {C.}~\bibnamefont {Dorrer}}, \bibinfo {author} {\bibfnamefont {J.~P.}\
  \bibnamefont {Palastro}},\ and\ \bibinfo {author} {\bibfnamefont
  {D.}~\bibnamefont {Froula}},\ }\href {https://doi.org/10.1364/OE.506112}
  {\bibfield  {journal} {\bibinfo  {journal} {Opt. Express}\ }\textbf {\bibinfo
  {volume} {32}},\ \bibinfo {pages} {576} (\bibinfo {year} {2024})}\BibitemShut
  {NoStop}%
\bibitem [{\citenamefont {Li}\ \emph {et~al.}(2024)\citenamefont {Li},
  \citenamefont {Miller}, \citenamefont {Pierce}, \citenamefont {Mori},
  \citenamefont {Thomas},\ and\ \citenamefont {Palastro}}]{Li_2024}%
  \BibitemOpen
  \bibfield  {author} {\bibinfo {author} {\bibfnamefont {D.}~\bibnamefont
  {Li}}, \bibinfo {author} {\bibfnamefont {K.~G.}\ \bibnamefont {Miller}},
  \bibinfo {author} {\bibfnamefont {J.~R.}\ \bibnamefont {Pierce}}, \bibinfo
  {author} {\bibfnamefont {W.~B.}\ \bibnamefont {Mori}}, \bibinfo {author}
  {\bibfnamefont {A.~G.~R.}\ \bibnamefont {Thomas}},\ and\ \bibinfo {author}
  {\bibfnamefont {J.~P.}\ \bibnamefont {Palastro}},\ }\href@noop {} {\bibfield
  {journal} {\bibinfo  {journal} {Phys. Rev. Res.}\ }\textbf {\bibinfo {volume}
  {6}},\ \bibinfo {pages} {013272} (\bibinfo {year} {2024})}\BibitemShut
  {NoStop}%
\bibitem [{\citenamefont {Di~Piazza}(2021)}]{DiPiazza:2020wxp}%
  \BibitemOpen
  \bibfield  {author} {\bibinfo {author} {\bibfnamefont {A.}~\bibnamefont
  {Di~Piazza}},\ }\href {https://doi.org/10.1103/PhysRevA.103.012215}
  {\bibfield  {journal} {\bibinfo  {journal} {Phys. Rev. A}\ }\textbf {\bibinfo
  {volume} {103}},\ \bibinfo {pages} {012215} (\bibinfo {year}
  {2021})}\BibitemShut {NoStop}%
\bibitem [{\citenamefont {Formanek}\ \emph {et~al.}(2024)\citenamefont
  {Formanek}, \citenamefont {Palastro}, \citenamefont {Ramsey}, \citenamefont
  {Weber},\ and\ \citenamefont {Di~Piazza}}]{Formanek:2023mkx}%
  \BibitemOpen
  \bibfield  {author} {\bibinfo {author} {\bibfnamefont {M.}~\bibnamefont
  {Formanek}}, \bibinfo {author} {\bibfnamefont {J.~P.}\ \bibnamefont
  {Palastro}}, \bibinfo {author} {\bibfnamefont {D.}~\bibnamefont {Ramsey}},
  \bibinfo {author} {\bibfnamefont {S.}~\bibnamefont {Weber}},\ and\ \bibinfo
  {author} {\bibfnamefont {A.}~\bibnamefont {Di~Piazza}},\ }\href
  {https://doi.org/10.1103/PhysRevD.109.056009} {\bibfield  {journal} {\bibinfo
   {journal} {Phys. Rev. D}\ }\textbf {\bibinfo {volume} {109}},\ \bibinfo
  {pages} {056009} (\bibinfo {year} {2024})}\BibitemShut {NoStop}%
\bibitem [{\citenamefont {Formanek}\ \emph {et~al.}(2025)\citenamefont
  {Formanek}, \citenamefont {Palastro}, \citenamefont {Ramsey},\ and\
  \citenamefont {Di~Piazza}}]{Formanek2025}%
  \BibitemOpen
  \bibfield  {author} {\bibinfo {author} {\bibfnamefont {M.}~\bibnamefont
  {Formanek}}, \bibinfo {author} {\bibfnamefont {J.~P.}\ \bibnamefont
  {Palastro}}, \bibinfo {author} {\bibfnamefont {D.}~\bibnamefont {Ramsey}},\
  and\ \bibinfo {author} {\bibfnamefont {A.}~\bibnamefont {Di~Piazza}},\ }\href
  {https://doi.org/10.1103/hlcs-wwyb} {\bibfield  {journal} {\bibinfo
  {journal} {Phys. Rev. A}\ }\textbf {\bibinfo {volume} {112}},\ \bibinfo
  {pages} {L051102} (\bibinfo {year} {2025})}\BibitemShut {NoStop}%
\bibitem [{\citenamefont {Adamo}\ and\ \citenamefont
  {Ilderton}(2025)}]{Adamo_2025}%
  \BibitemOpen
  \bibfield  {author} {\bibinfo {author} {\bibfnamefont {T.}~\bibnamefont
  {Adamo}}\ and\ \bibinfo {author} {\bibfnamefont {A.}~\bibnamefont
  {Ilderton}},\ }\href@noop {} {\bibfield  {journal} {\bibinfo  {journal}
  {Phys. Rev. D}\ }\textbf {\bibinfo {volume} {111}},\ \bibinfo {pages}
  {125005} (\bibinfo {year} {2025})}\BibitemShut {NoStop}%
\bibitem [{\citenamefont {Liberman}\ \emph
  {et~al.}(2026{\natexlab{a}})\citenamefont {Liberman}, \citenamefont
  {Golovanov}, \citenamefont {Smartsev}, \citenamefont {Talposi}, \citenamefont
  {Tata},\ and\ \citenamefont {Malka}}]{Liberman2026PRR}%
  \BibitemOpen
  \bibfield  {author} {\bibinfo {author} {\bibfnamefont {A.}~\bibnamefont
  {Liberman}}, \bibinfo {author} {\bibfnamefont {A.}~\bibnamefont {Golovanov}},
  \bibinfo {author} {\bibfnamefont {S.}~\bibnamefont {Smartsev}}, \bibinfo
  {author} {\bibfnamefont {A.-M.}\ \bibnamefont {Talposi}}, \bibinfo {author}
  {\bibfnamefont {S.}~\bibnamefont {Tata}},\ and\ \bibinfo {author}
  {\bibfnamefont {V.}~\bibnamefont {Malka}},\ }\href
  {https://doi.org/10.1103/x5pr-mdvj} {\bibfield  {journal} {\bibinfo
  {journal} {Phys. Rev. Res.}\ }\textbf {\bibinfo {volume} {8}},\ \bibinfo
  {pages} {L022001} (\bibinfo {year} {2026}{\natexlab{a}})}\BibitemShut
  {NoStop}%
\bibitem [{\citenamefont {Liberman}\ \emph
  {et~al.}(2026{\natexlab{b}})\citenamefont {Liberman}, \citenamefont
  {Golovanov}, \citenamefont {Smartsev}, \citenamefont {Talposi}, \citenamefont
  {Tata},\ and\ \citenamefont {Malka}}]{Liberman2026FFLWFA}%
  \BibitemOpen
  \bibfield  {author} {\bibinfo {author} {\bibfnamefont {A.}~\bibnamefont
  {Liberman}}, \bibinfo {author} {\bibfnamefont {A.}~\bibnamefont {Golovanov}},
  \bibinfo {author} {\bibfnamefont {S.}~\bibnamefont {Smartsev}}, \bibinfo
  {author} {\bibfnamefont {A.-M.}\ \bibnamefont {Talposi}}, \bibinfo {author}
  {\bibfnamefont {S.}~\bibnamefont {Tata}},\ and\ \bibinfo {author}
  {\bibfnamefont {V.}~\bibnamefont {Malka}},\ }\href
  {https://doi.org/10.48550/arXiv.2604.05771} {\bibinfo {title} {Electron
  acceleration in a flying-focus laser wakefield accelerator}} (\bibinfo {year}
  {2026}{\natexlab{b}}),\ \Eprint {https://arxiv.org/abs/2604.05771}
  {arXiv:2604.05771} \BibitemShut {NoStop}%
\bibitem [{\citenamefont {Arrowsmith}\ \emph {et~al.}(2026)\citenamefont
  {Arrowsmith}, \citenamefont {Miller}, \citenamefont {Ambat}, \citenamefont
  {Bahk}, \citenamefont {Begishev}, \citenamefont {Bromage}, \citenamefont
  {Bucht}, \citenamefont {Dauphin}, \citenamefont {Dorrer}, \citenamefont
  {Jeon}, \citenamefont {Kendrick}, \citenamefont {LaBelle}, \citenamefont
  {Mack}, \citenamefont {Martin}, \citenamefont {Mileham}, \citenamefont {Qin},
  \citenamefont {Pigeon}, \citenamefont {Raymond}, \citenamefont {Romanofsky},
  \citenamefont {Rinderknecht}, \citenamefont {Roides}, \citenamefont
  {Szczepanski}, \citenamefont {Settle}, \citenamefont {Spilatro},
  \citenamefont {Webb}, \citenamefont {Shaw}, \citenamefont {Palastro},\ and\
  \citenamefont {Froula}}]{Arrowsmith_2026}%
  \BibitemOpen
  \bibfield  {author} {\bibinfo {author} {\bibfnamefont {C.~D.}\ \bibnamefont
  {Arrowsmith}}, \bibinfo {author} {\bibfnamefont {K.~G.}\ \bibnamefont
  {Miller}}, \bibinfo {author} {\bibfnamefont {M.~V.}\ \bibnamefont {Ambat}},
  \bibinfo {author} {\bibfnamefont {S.-W.}\ \bibnamefont {Bahk}}, \bibinfo
  {author} {\bibfnamefont {I.~A.}\ \bibnamefont {Begishev}}, \bibinfo {author}
  {\bibfnamefont {J.}~\bibnamefont {Bromage}}, \bibinfo {author} {\bibfnamefont
  {S.}~\bibnamefont {Bucht}}, \bibinfo {author} {\bibfnamefont
  {N.}~\bibnamefont {Dauphin}}, \bibinfo {author} {\bibfnamefont
  {C.}~\bibnamefont {Dorrer}}, \bibinfo {author} {\bibfnamefont
  {C.}~\bibnamefont {Jeon}}, \bibinfo {author} {\bibfnamefont {J.}~\bibnamefont
  {Kendrick}}, \bibinfo {author} {\bibfnamefont {I.~A.}\ \bibnamefont
  {LaBelle}}, \bibinfo {author} {\bibfnamefont {L.~S.}\ \bibnamefont {Mack}},
  \bibinfo {author} {\bibfnamefont {A.~L.}\ \bibnamefont {Martin}}, \bibinfo
  {author} {\bibfnamefont {C.}~\bibnamefont {Mileham}}, \bibinfo {author}
  {\bibfnamefont {S.}~\bibnamefont {Qin}}, \bibinfo {author} {\bibfnamefont
  {J.~J.}\ \bibnamefont {Pigeon}}, \bibinfo {author} {\bibfnamefont
  {A.}~\bibnamefont {Raymond}}, \bibinfo {author} {\bibfnamefont
  {M.}~\bibnamefont {Romanofsky}}, \bibinfo {author} {\bibfnamefont {H.~G.}\
  \bibnamefont {Rinderknecht}}, \bibinfo {author} {\bibfnamefont {R.~G.}\
  \bibnamefont {Roides}}, \bibinfo {author} {\bibfnamefont {J.}~\bibnamefont
  {Szczepanski}}, \bibinfo {author} {\bibfnamefont {I.~A.}\ \bibnamefont
  {Settle}}, \bibinfo {author} {\bibfnamefont {M.}~\bibnamefont {Spilatro}},
  \bibinfo {author} {\bibfnamefont {B.}~\bibnamefont {Webb}}, \bibinfo {author}
  {\bibfnamefont {J.~L.}\ \bibnamefont {Shaw}}, \bibinfo {author}
  {\bibfnamefont {J.~P.}\ \bibnamefont {Palastro}},\ and\ \bibinfo {author}
  {\bibfnamefont {D.~H.}\ \bibnamefont {Froula}},\ }\bibfield  {journal}
  {\bibinfo  {journal} {Nature Phys.}\ }\href
  {https://doi.org/10.1038/s41567-026-03352-x} {10.1038/s41567-026-03352-x}
  (\bibinfo {year} {2026})\BibitemShut {NoStop}%
\bibitem [{\citenamefont {Kabacinski}\ \emph {et~al.}(2023)\citenamefont
  {Kabacinski}, \citenamefont {Oliva}, \citenamefont {Tissandier},
  \citenamefont {Gautier}, \citenamefont {Kozlov{\'a}}, \citenamefont {Goddet},
  \citenamefont {Andriyash}, \citenamefont {Thaury}, \citenamefont {Zeitoun},\
  and\ \citenamefont {Sebban}}]{Kabacinski_2023}%
  \BibitemOpen
  \bibfield  {author} {\bibinfo {author} {\bibfnamefont {A.}~\bibnamefont
  {Kabacinski}}, \bibinfo {author} {\bibfnamefont {E.}~\bibnamefont {Oliva}},
  \bibinfo {author} {\bibfnamefont {F.}~\bibnamefont {Tissandier}}, \bibinfo
  {author} {\bibfnamefont {J.}~\bibnamefont {Gautier}}, \bibinfo {author}
  {\bibfnamefont {M.}~\bibnamefont {Kozlov{\'a}}}, \bibinfo {author}
  {\bibfnamefont {J.-P.}\ \bibnamefont {Goddet}}, \bibinfo {author}
  {\bibfnamefont {I.~A.}\ \bibnamefont {Andriyash}}, \bibinfo {author}
  {\bibfnamefont {C.}~\bibnamefont {Thaury}}, \bibinfo {author} {\bibfnamefont
  {P.}~\bibnamefont {Zeitoun}},\ and\ \bibinfo {author} {\bibfnamefont
  {S.}~\bibnamefont {Sebban}},\ }\href@noop {} {\bibfield  {journal} {\bibinfo
  {journal} {Nature Photon.}\ }\textbf {\bibinfo {volume} {17}},\ \bibinfo
  {pages} {354} (\bibinfo {year} {2023})}\BibitemShut {NoStop}%
\bibitem [{LEP()}]{LEP}%
  \BibitemOpen
  \href@noop {} {\bibinfo {title} {{Large Electron-Positron (LEP) collider}}},\
  \bibinfo {howpublished}
  {\url{https://home.cern/science/accelerators/large-electron-positron-collider/}}\BibitemShut
  {NoStop}%
\bibitem [{\citenamefont {Yakimenko}\ \emph {et~al.}(2019)\citenamefont
  {Yakimenko}, \citenamefont {Alsberg}, \citenamefont {Bong}, \citenamefont
  {Bouchard}, \citenamefont {Clarke}, \citenamefont {Emma}, \citenamefont
  {Green}, \citenamefont {Hast}, \citenamefont {Hogan}, \citenamefont {Seabury}
  \emph {et~al.}}]{Facet-II}%
  \BibitemOpen
  \bibfield  {author} {\bibinfo {author} {\bibfnamefont {V.}~\bibnamefont
  {Yakimenko}}, \bibinfo {author} {\bibfnamefont {L.}~\bibnamefont {Alsberg}},
  \bibinfo {author} {\bibfnamefont {E.}~\bibnamefont {Bong}}, \bibinfo {author}
  {\bibfnamefont {G.}~\bibnamefont {Bouchard}}, \bibinfo {author}
  {\bibfnamefont {C.}~\bibnamefont {Clarke}}, \bibinfo {author} {\bibfnamefont
  {C.}~\bibnamefont {Emma}}, \bibinfo {author} {\bibfnamefont {S.}~\bibnamefont
  {Green}}, \bibinfo {author} {\bibfnamefont {C.}~\bibnamefont {Hast}},
  \bibinfo {author} {\bibfnamefont {M.~J.}\ \bibnamefont {Hogan}}, \bibinfo
  {author} {\bibfnamefont {J.}~\bibnamefont {Seabury}}, \emph {et~al.},\ }\href
  {https://doi.org/10.1103/PhysRevAccelBeams.22.101301} {\bibfield  {journal}
  {\bibinfo  {journal} {Phys. Rev. Accel. Beams}\ }\textbf {\bibinfo {volume}
  {22}},\ \bibinfo {pages} {101301} (\bibinfo {year} {2019})}\BibitemShut
  {NoStop}%
\bibitem [{\citenamefont {Abramowicz}\ \emph {et~al.}(2024)\citenamefont
  {Abramowicz} \emph {et~al.}}]{LUXE:2024TDR}%
  \BibitemOpen
  \bibfield  {author} {\bibinfo {author} {\bibfnamefont {H.}~\bibnamefont
  {Abramowicz}} \emph {et~al.} (\bibinfo {collaboration} {LUXE
  Collaboration}),\ }\bibfield  {journal} {\bibinfo  {journal} {Eur. Phys. J.
  ST}\ }\href {https://doi.org/10.1140/epjs/s11734-024-01164-9}
  {10.1140/epjs/s11734-024-01164-9} (\bibinfo {year} {2024}),\ \Eprint
  {https://arxiv.org/abs/2308.00515} {arXiv:2308.00515 [physics.ins-det]}
  \BibitemShut {NoStop}%
\bibitem [{\citenamefont {Picksley}\ \emph {et~al.}(2024)\citenamefont
  {Picksley}, \citenamefont {Stackhouse}, \citenamefont {Benedetti},
  \citenamefont {Nakamura}, \citenamefont {Tsai}, \citenamefont {Li},
  \citenamefont {Miao}, \citenamefont {Shrock}, \citenamefont {Rockafellow},
  \citenamefont {Milchberg}, \citenamefont {Schroeder}, \citenamefont {van
  Tilborg}, \citenamefont {Esarey}, \citenamefont {Geddes},\ and\ \citenamefont
  {Gonsalves}}]{Picksley_2024}%
  \BibitemOpen
  \bibfield  {author} {\bibinfo {author} {\bibfnamefont {A.}~\bibnamefont
  {Picksley}}, \bibinfo {author} {\bibfnamefont {J.}~\bibnamefont
  {Stackhouse}}, \bibinfo {author} {\bibfnamefont {C.}~\bibnamefont
  {Benedetti}}, \bibinfo {author} {\bibfnamefont {K.}~\bibnamefont {Nakamura}},
  \bibinfo {author} {\bibfnamefont {H.~E.}\ \bibnamefont {Tsai}}, \bibinfo
  {author} {\bibfnamefont {R.}~\bibnamefont {Li}}, \bibinfo {author}
  {\bibfnamefont {B.}~\bibnamefont {Miao}}, \bibinfo {author} {\bibfnamefont
  {J.~E.}\ \bibnamefont {Shrock}}, \bibinfo {author} {\bibfnamefont
  {E.}~\bibnamefont {Rockafellow}}, \bibinfo {author} {\bibfnamefont {H.~M.}\
  \bibnamefont {Milchberg}}, \bibinfo {author} {\bibfnamefont {C.~B.}\
  \bibnamefont {Schroeder}}, \bibinfo {author} {\bibfnamefont {J.}~\bibnamefont
  {van Tilborg}}, \bibinfo {author} {\bibfnamefont {E.}~\bibnamefont {Esarey}},
  \bibinfo {author} {\bibfnamefont {C.~G.~R.}\ \bibnamefont {Geddes}},\ and\
  \bibinfo {author} {\bibfnamefont {A.~J.}\ \bibnamefont {Gonsalves}},\
  }\href@noop {} {\bibfield  {journal} {\bibinfo  {journal} {Phys. Rev. Lett.}\
  }\textbf {\bibinfo {volume} {133}},\ \bibinfo {pages} {255001} (\bibinfo
  {year} {2024})}\BibitemShut {NoStop}%
\bibitem [{\citenamefont {Shaw}\ \emph {et~al.}(2025)\citenamefont {Shaw},
  \citenamefont {Ambat}, \citenamefont {Miller}, \citenamefont {Boni},
  \citenamefont {LaBelle}, \citenamefont {Mori}, \citenamefont {Pigeon},
  \citenamefont {Rigatti}, \citenamefont {Settle}, \citenamefont {Mack} \emph
  {et~al.}}]{Shaw2025OPAL100GeV}%
  \BibitemOpen
  \bibfield  {author} {\bibinfo {author} {\bibfnamefont {J.~L.}\ \bibnamefont
  {Shaw}}, \bibinfo {author} {\bibfnamefont {M.~V.}\ \bibnamefont {Ambat}},
  \bibinfo {author} {\bibfnamefont {K.~G.}\ \bibnamefont {Miller}}, \bibinfo
  {author} {\bibfnamefont {R.}~\bibnamefont {Boni}}, \bibinfo {author}
  {\bibfnamefont {I.~A.}\ \bibnamefont {LaBelle}}, \bibinfo {author}
  {\bibfnamefont {W.~B.}\ \bibnamefont {Mori}}, \bibinfo {author}
  {\bibfnamefont {J.~J.}\ \bibnamefont {Pigeon}}, \bibinfo {author}
  {\bibfnamefont {A.}~\bibnamefont {Rigatti}}, \bibinfo {author} {\bibfnamefont
  {I.~A.}\ \bibnamefont {Settle}}, \bibinfo {author} {\bibfnamefont {L.~S.}\
  \bibnamefont {Mack}}, \emph {et~al.},\ }\href
  {https://doi.org/10.1063/5.0274780} {\bibfield  {journal} {\bibinfo
  {journal} {Physics of Plasmas}\ }\textbf {\bibinfo {volume} {32}},\ \bibinfo
  {pages} {083107} (\bibinfo {year} {2025})}\BibitemShut {NoStop}%
\bibitem [{FCC()}]{FCC}%
  \BibitemOpen
  \href@noop {} {\bibinfo {title} {{Future Circular Collider (FCC)}}},\
  \bibinfo {howpublished}
  {\url{https://home.cern/science/accelerators/future-circular-collider/}}\BibitemShut
  {NoStop}%
\bibitem [{CLI()}]{CLIC}%
  \BibitemOpen
  \href@noop {} {\bibinfo {title} {{Compact Linear Collider (CLIC)}}},\
  \bibinfo {howpublished}
  {\url{https://home.cern/science/accelerators/compact-linear-collider/}}\BibitemShut
  {NoStop}%
\bibitem [{ILC()}]{ILC}%
  \BibitemOpen
  \href@noop {} {\bibinfo {title} {{International Linear Collider (ILC)}}},\
  \bibinfo {howpublished} {\url{https://linearcollider.org/}}\BibitemShut
  {NoStop}%
\bibitem [{\citenamefont {Burton}\ and\ \citenamefont
  {Noble}(2014)}]{Burton:2014wsa}%
  \BibitemOpen
  \bibfield  {author} {\bibinfo {author} {\bibfnamefont {D.~A.}\ \bibnamefont
  {Burton}}\ and\ \bibinfo {author} {\bibfnamefont {A.}~\bibnamefont {Noble}},\
  }\href {https://doi.org/10.1080/00107514.2014.886840} {\bibfield  {journal}
  {\bibinfo  {journal} {Contemp. Phys.}\ }\textbf {\bibinfo {volume} {55}},\
  \bibinfo {pages} {110} (\bibinfo {year} {2014})}\BibitemShut {NoStop}%
\bibitem [{SM()}]{SM}%
  \BibitemOpen
  \href@noop {} {\bibinfo  {journal} {See Supplemental Material at {\it url
  will be inserted by publisher} for simulation details}\ }\BibitemShut
  {NoStop}%
\bibitem [{\citenamefont {Blackburn}\ \emph {et~al.}(2023)\citenamefont
  {Blackburn}, \citenamefont {King},\ and\ \citenamefont
  {Tang}}]{Blackburn2023}%
  \BibitemOpen
\bibfield  {journal} {  }\bibfield  {author} {\bibinfo {author} {\bibfnamefont
  {T.~G.}\ \bibnamefont {Blackburn}}, \bibinfo {author} {\bibfnamefont
  {B.}~\bibnamefont {King}},\ and\ \bibinfo {author} {\bibfnamefont
  {S.}~\bibnamefont {Tang}},\ }\href {https://doi.org/10.1063/5.0159963}
  {\bibfield  {journal} {\bibinfo  {journal} {Phys. Plasmas}\ }\textbf
  {\bibinfo {volume} {30}},\ \bibinfo {pages} {093903} (\bibinfo {year}
  {2023})}\BibitemShut {NoStop}%
\bibitem [{\citenamefont {Reiss}(1962)}]{Reiss:1962nhe}%
  \BibitemOpen
  \bibfield  {author} {\bibinfo {author} {\bibfnamefont {H.~R.}\ \bibnamefont
  {Reiss}},\ }\href {https://doi.org/10.1063/1.1703787} {\bibfield  {journal}
  {\bibinfo  {journal} {J. Math. Phys.}\ }\textbf {\bibinfo {volume} {3}},\
  \bibinfo {pages} {59} (\bibinfo {year} {1962})}\BibitemShut {NoStop}%
\bibitem [{\citenamefont {Di~Piazza}\ \emph
  {et~al.}(2019{\natexlab{a}})\citenamefont {Di~Piazza}, \citenamefont
  {Tamburini}, \citenamefont {Meuren},\ and\ \citenamefont
  {Keitel}}]{DiPiazza:2018bfu}%
  \BibitemOpen
  \bibfield  {author} {\bibinfo {author} {\bibfnamefont {A.}~\bibnamefont
  {Di~Piazza}}, \bibinfo {author} {\bibfnamefont {M.}~\bibnamefont
  {Tamburini}}, \bibinfo {author} {\bibfnamefont {S.}~\bibnamefont {Meuren}},\
  and\ \bibinfo {author} {\bibfnamefont {C.~H.}\ \bibnamefont {Keitel}},\
  }\href {https://doi.org/10.1103/PhysRevA.99.022125} {\bibfield  {journal}
  {\bibinfo  {journal} {Phys. Rev. A}\ }\textbf {\bibinfo {volume} {99}},\
  \bibinfo {pages} {022125} (\bibinfo {year} {2019}{\natexlab{a}})}\BibitemShut
  {NoStop}%
\bibitem [{\citenamefont {Ilderton}\ \emph {et~al.}(2019)\citenamefont
  {Ilderton}, \citenamefont {King},\ and\ \citenamefont
  {Seipt}}]{Ilderton_2019_b}%
  \BibitemOpen
  \bibfield  {author} {\bibinfo {author} {\bibfnamefont {A.}~\bibnamefont
  {Ilderton}}, \bibinfo {author} {\bibfnamefont {B.}~\bibnamefont {King}},\
  and\ \bibinfo {author} {\bibfnamefont {D.}~\bibnamefont {Seipt}},\ }\href
  {https://doi.org/10.1103/PhysRevA.99.042121} {\bibfield  {journal} {\bibinfo
  {journal} {Phys. Rev. A}\ }\textbf {\bibinfo {volume} {99}},\ \bibinfo
  {pages} {042121} (\bibinfo {year} {2019})}\BibitemShut {NoStop}%
\bibitem [{\citenamefont {Di~Piazza}\ \emph
  {et~al.}(2019{\natexlab{b}})\citenamefont {Di~Piazza}, \citenamefont
  {Tamburini}, \citenamefont {Meuren},\ and\ \citenamefont
  {Keitel}}]{Di_Piazza_2019}%
  \BibitemOpen
  \bibfield  {author} {\bibinfo {author} {\bibfnamefont {A.}~\bibnamefont
  {Di~Piazza}}, \bibinfo {author} {\bibfnamefont {M.}~\bibnamefont
  {Tamburini}}, \bibinfo {author} {\bibfnamefont {S.}~\bibnamefont {Meuren}},\
  and\ \bibinfo {author} {\bibfnamefont {C.~H.}\ \bibnamefont {Keitel}},\
  }\href {https://doi.org/10.1103/PhysRevA.99.022125} {\bibfield  {journal}
  {\bibinfo  {journal} {Phys. Rev. A}\ }\textbf {\bibinfo {volume} {99}},\
  \bibinfo {pages} {022125} (\bibinfo {year} {2019}{\natexlab{b}})}\BibitemShut
  {NoStop}%
\bibitem [{\citenamefont {Salamin}(2007)}]{Salamin_2007}%
  \BibitemOpen
  \bibfield  {author} {\bibinfo {author} {\bibfnamefont {Y.~I.}\ \bibnamefont
  {Salamin}},\ }\href@noop {} {\bibfield  {journal} {\bibinfo  {journal} {Appl.
  Phys. B}\ }\textbf {\bibinfo {volume} {86}},\ \bibinfo {pages} {319}
  (\bibinfo {year} {2007})}\BibitemShut {NoStop}%
\bibitem [{\citenamefont {Di~Piazza}\ \emph {et~al.}(2010)\citenamefont
  {Di~Piazza}, \citenamefont {Hatsagortsyan},\ and\ \citenamefont
  {Keitel}}]{Di_Piazza_2010}%
  \BibitemOpen
  \bibfield  {author} {\bibinfo {author} {\bibfnamefont {A.}~\bibnamefont
  {Di~Piazza}}, \bibinfo {author} {\bibfnamefont {K.~Z.}\ \bibnamefont
  {Hatsagortsyan}},\ and\ \bibinfo {author} {\bibfnamefont {C.~H.}\
  \bibnamefont {Keitel}},\ }\href
  {https://doi.org/10.1103/PhysRevLett.105.220403} {\bibfield  {journal}
  {\bibinfo  {journal} {Phys. Rev. Lett.}\ }\textbf {\bibinfo {volume} {105}},\
  \bibinfo {pages} {220403} (\bibinfo {year} {2010})}\BibitemShut {NoStop}%
\bibitem [{\citenamefont {Altarelli}\ \emph {et~al.}(2007)\citenamefont
  {Altarelli} \emph {et~al.}}]{XFEL2007}%
  \BibitemOpen
  \bibfield  {author} {\bibinfo {author} {\bibfnamefont {M.}~\bibnamefont
  {Altarelli}} \emph {et~al.},\ }\href@noop {} {\emph {\bibinfo {title} {XFEL:
  The European X-Ray Free-Electron Laser Technical Design Report}}},\ \bibinfo
  {type} {Tech. Rep.}\ \bibinfo {number} {DESY 2006-097}\ (\bibinfo
  {institution} {DESY},\ \bibinfo {address} {Hamburg},\ \bibinfo {year}
  {2007})\BibitemShut {NoStop}%
\bibitem [{\citenamefont {Palmer}\ \emph {et~al.}(2025)\citenamefont {Palmer},
  \citenamefont {Gonz\'{a}lez-D\'{i}az}, \citenamefont {Eichner}, \citenamefont
  {Braun}, \citenamefont {Jiang}, \citenamefont {H\"{u}lsenbusch},
  \citenamefont {Werle}, \citenamefont {Winkelmann}, \citenamefont {Vidoli},
  \citenamefont {Yousefi} \emph {et~al.}}]{Palmer:25}%
  \BibitemOpen
  \bibfield  {author} {\bibinfo {author} {\bibfnamefont {G.}~\bibnamefont
  {Palmer}}, \bibinfo {author} {\bibfnamefont {J.~B.}\ \bibnamefont
  {Gonz\'{a}lez-D\'{i}az}}, \bibinfo {author} {\bibfnamefont {T.}~\bibnamefont
  {Eichner}}, \bibinfo {author} {\bibfnamefont {C.}~\bibnamefont {Braun}},
  \bibinfo {author} {\bibfnamefont {M.}~\bibnamefont {Jiang}}, \bibinfo
  {author} {\bibfnamefont {T.}~\bibnamefont {H\"{u}lsenbusch}}, \bibinfo
  {author} {\bibfnamefont {C.}~\bibnamefont {Werle}}, \bibinfo {author}
  {\bibfnamefont {L.}~\bibnamefont {Winkelmann}}, \bibinfo {author}
  {\bibfnamefont {C.}~\bibnamefont {Vidoli}}, \bibinfo {author} {\bibfnamefont
  {A.}~\bibnamefont {Yousefi}}, \emph {et~al.},\ }in\ \href
  {https://opg.optica.org/abstract.cfm?URI=CLEO_Europe-2025-cg_4_1} {\emph
  {\bibinfo {booktitle} {Conference on Lasers and Electro-Optics/Europe
  (CLEO/Europe 2025) and European Quantum Electronics Conference (EQEC
  2025)}}}\ (\bibinfo  {publisher} {Optica Publishing Group},\ \bibinfo {year}
  {2025})\ p.\ \bibinfo {pages} {cg\_4\_1}\BibitemShut {NoStop}%
\bibitem [{KBE(2017)}]{KBELLA2017}%
  \BibitemOpen
  \href
  {https://www2.lbl.gov/LBL-Programs/atap/Report_Workshop_k-BELLA_laser_tech_final.pdf}
  {\emph {\bibinfo {title} {Report of Workshop on Laser Technology for k-BELLA
  and Beyond}}},\ \bibinfo {type} {Tech. Rep.}\ \bibinfo {number} {17-AF-4192}\
  (\bibinfo  {institution} {Lawrence Berkeley National Laboratory},\ \bibinfo
  {year} {2017})\BibitemShut {NoStop}%
\bibitem [{\citenamefont {Kiani}\ \emph {et~al.}(2023)\citenamefont {Kiani},
  \citenamefont {Zhou}, \citenamefont {Bahk}, \citenamefont {Bromage},
  \citenamefont {Bruhwiler}, \citenamefont {Campbell}, \citenamefont {Chang},
  \citenamefont {Chowdhury}, \citenamefont {Downer}, \citenamefont {Du} \emph
  {et~al.}}]{Kiani2023}%
  \BibitemOpen
  \bibfield  {author} {\bibinfo {author} {\bibfnamefont {L.}~\bibnamefont
  {Kiani}}, \bibinfo {author} {\bibfnamefont {T.}~\bibnamefont {Zhou}},
  \bibinfo {author} {\bibfnamefont {S.-W.}\ \bibnamefont {Bahk}}, \bibinfo
  {author} {\bibfnamefont {J.}~\bibnamefont {Bromage}}, \bibinfo {author}
  {\bibfnamefont {D.}~\bibnamefont {Bruhwiler}}, \bibinfo {author}
  {\bibfnamefont {E.~M.}\ \bibnamefont {Campbell}}, \bibinfo {author}
  {\bibfnamefont {Z.}~\bibnamefont {Chang}}, \bibinfo {author} {\bibfnamefont
  {E.}~\bibnamefont {Chowdhury}}, \bibinfo {author} {\bibfnamefont
  {M.}~\bibnamefont {Downer}}, \bibinfo {author} {\bibfnamefont
  {Q.}~\bibnamefont {Du}}, \emph {et~al.},\ }\href
  {https://doi.org/10.1088/1748-0221/18/08/T08006} {\bibfield  {journal}
  {\bibinfo  {journal} {Journal of Instrumentation}\ }\textbf {\bibinfo
  {volume} {18}},\ \bibinfo {pages} {T08006}}\BibitemShut {NoStop}%
\bibitem [{\citenamefont {Formanek}\ \emph {et~al.}(2022)\citenamefont
  {Formanek}, \citenamefont {Ramsey}, \citenamefont {Palastro},\ and\
  \citenamefont {Di~Piazza}}]{Formanek:2021bpw}%
  \BibitemOpen
  \bibfield  {author} {\bibinfo {author} {\bibfnamefont {M.}~\bibnamefont
  {Formanek}}, \bibinfo {author} {\bibfnamefont {D.}~\bibnamefont {Ramsey}},
  \bibinfo {author} {\bibfnamefont {J.~P.}\ \bibnamefont {Palastro}},\ and\
  \bibinfo {author} {\bibfnamefont {A.}~\bibnamefont {Di~Piazza}},\ }\href
  {https://doi.org/10.1103/PhysRevA.105.L020203} {\bibfield  {journal}
  {\bibinfo  {journal} {Phys. Rev. A}\ }\textbf {\bibinfo {volume} {105}},\
  \bibinfo {pages} {L020203} (\bibinfo {year} {2022})}\BibitemShut {NoStop}%
\bibitem [{\citenamefont {Ramsey}\ \emph {et~al.}(2023)\citenamefont {Ramsey},
  \citenamefont {Di~Piazza}, \citenamefont {Formanek}, \citenamefont {Franke},
  \citenamefont {Froula}, \citenamefont {Malaca}, \citenamefont {Mori},
  \citenamefont {Pierce}, \citenamefont {Simpson}, \citenamefont {Vieira} \emph
  {et~al.}}]{ramsey2023exact}%
  \BibitemOpen
  \bibfield  {author} {\bibinfo {author} {\bibfnamefont {D.}~\bibnamefont
  {Ramsey}}, \bibinfo {author} {\bibfnamefont {A.}~\bibnamefont {Di~Piazza}},
  \bibinfo {author} {\bibfnamefont {M.}~\bibnamefont {Formanek}}, \bibinfo
  {author} {\bibfnamefont {P.}~\bibnamefont {Franke}}, \bibinfo {author}
  {\bibfnamefont {D.~H.}\ \bibnamefont {Froula}}, \bibinfo {author}
  {\bibfnamefont {B.}~\bibnamefont {Malaca}}, \bibinfo {author} {\bibfnamefont
  {W.~B.}\ \bibnamefont {Mori}}, \bibinfo {author} {\bibfnamefont {J.~R.}\
  \bibnamefont {Pierce}}, \bibinfo {author} {\bibfnamefont {T.~T.}\
  \bibnamefont {Simpson}}, \bibinfo {author} {\bibfnamefont {J.}~\bibnamefont
  {Vieira}}, \emph {et~al.},\ }\href
  {https://doi.org/10.1103/PhysRevA.107.013513} {\bibfield  {journal} {\bibinfo
   {journal} {Phys. Rev. A}\ }\textbf {\bibinfo {volume} {107}},\ \bibinfo
  {pages} {013513} (\bibinfo {year} {2023})}\BibitemShut {NoStop}%
\end{thebibliography}

\end{document}